\documentclass[11pt,a4paper]{article}
\pdfoutput=1
\usepackage{jcappub}

\usepackage{marginnote}
\usepackage{placeins}
\usepackage[normalem]{ulem}
\usepackage{epsfig}
\usepackage{bm}
\usepackage{tabu}
\usepackage{boldline}
\usepackage{xspace}
\usepackage{slashed}
\usepackage{multirow}
\usepackage{diagbox}
\usepackage[title]{appendix}
\usepackage{subcaption}
\usepackage[table]{xcolor}
\usepackage{enumitem}
\usepackage[utf8]{inputenc}
\usepackage{colortbl}
\usepackage{multirow}
\usepackage[normalem]{ulem}
\usepackage{setspace}
\usepackage{float}

\usepackage{cellspace}
\hypersetup{colorlinks,bookmarksopen,bookmarksnumbered,
linkcolor=cyan,urlcolor=black,citecolor=purple}

\newcommand{\vmin}{v_{\rm min}}
\newcommand{\vesc}{v_{\rm esc}}
\newcommand{\rhoDM}{\rho_{\rm DM}}

\newcommand{\sige}{\overline\sigma_e}

\definecolor{orange}{rgb}{1,0.5,0}

\definecolor{blue-violet}{rgb}{0.54, 0.17, 0.89}

\definecolor{rose}{rgb}{0.72, 0.43, 0.47}

\DeclareMathOperator\erf{erf}

\graphicspath{{figs/}}

\begin{document}

\title{Dependence of Dark Matter - Electron Scattering on the Galactic Dark Matter Velocity Distribution}
\author[a,b]{Aria Radick}
\emailAdd{aradick@uoregon.edu}
\author[b]{, Anna-Maria Taki}
\emailAdd{ataki@uoregon.edu}
\author[a,b]{, Tien-Tien Yu}
\emailAdd{tientien@uoregon.edu}
\affiliation[a]{Department of Physics, University of Oregon, Eugene, OR 97403, USA}
\affiliation[b]{Institute for Fundamental Science, University of Oregon, Eugene, OR 97403, USA}

\date{\today}

%============ START ============%

\abstract{The rate of dark matter-electron scattering depends on the underlying velocity distribution of the dark matter halo. Importantly, dark matter-electron scattering is particularly sensitive to the high-velocity tail, which differs significantly amongst the various dark matter halo models. In this work, we summarize the leading halo models and discuss the various parameters which enter them. We recommend updated values for these parameters based on recent studies and measurements. Furthermore, we quantify the dependence of the dark matter-electron scattering rate on the choice of halo model and parameters, and demonstrate how these choices propagate to the predicted cross-section limits. The rate is most sensitive to changes in the circular velocity $v_0$; in silicon targets, we find that the changes in the rate predictions can range from ${\cal O}(0.01\%)$ to ${\cal O}(100\%)$ for contact interactions and ${\cal O}(10\%)$ to ${\cal O}(100\%)$ for long-range interactions.
}

\maketitle

\section{Introduction}

Understanding the nature of dark matter (DM) is one of the most important outstanding questions in physics. As such, there has been significant effort from both theorists and experimentalists to uncover its identity and cosmological origin. One of the main search strategies for DM is direct detection (DD); the original proposal for this method focuses on the elastic scattering of DM on atomic nuclei within a detector, typically placed deep underground~\cite{Goodman:1984dc}, which produces a detectable nuclear recoil in a low-background material. This has led to an extensive experimental program that has resulted in strong constraints on the DM-nucleon cross section for DM masses above a few GeV, in materials ranging from noble liquid targets~\cite{Akerib:2016vxi,Cui:2017nnn,Agnes:2018ves,Aprile:2018dbl} to crystal targets~\cite{Agnese:2017njq,Abdelhameed:2019hmk,Armengaud:2019kfj}.
A new avenue of DD has emerged recently, which targets sub-GeV DM candidates by probing DM-electron interactions~\cite{Aguilar-Arevalo:2019wdi,Aprile:2019xxb,Amaral:2020ryn,Arnaud:2020svb,Barak:2020fql}.

The interpretations of DD experimental results in terms of DM-Standard Model interactions are sensitive to the astrophysical properties of the local DM halo distribution, such as the local DM density, the mean DM velocity, and DM velocity dispersion. The benchmark distribution is the Standard Halo Model (SHM) in which the DM particles are distributed in an isothermal sphere with an isotropic Maxwellian velocity distribution~\cite{drukier1986detecting}; this type of distribution follows from a system of collisionless particles and gives a convenient proxy for the velocity profile of the DM halo due to its simple analytic form (see {\it e.g.}~\cite{Bozorgnia:2016ogo}).
Despite its convenience, the SHM fails to provide an accurate description of how DM is distributed throughout the Galaxy (see detailed discussion in $\S$\ref{sec:SHM} and $\S$\ref{sub:sims}).
This motivates the study of alternative halo models for the DM velocity distribution function (VDF). Given that the DM VDF cannot be reliably extracted from direct measurements, other methods to infer the DM halo distribution include performing numerical cosmological simulations (see relevant discussion in $\S$\ref{sub:sims}), employing Eddington's formalism~\cite{eddington1916distribution,lacroix2018anatomy}, or using stellar populations that faithfully trace the DM component~\cite{herzog2018empirical,necib2019inferred}.

The effects of the uncertainties in the DM VDF on the DD observables have been studied in the past for DM-nuclear scattering interactions~\cite{kuhlen2010dark,mccabe2010astrophysical, green2017astrophysical,nunez2019dark}. In this work, we investigate the impact of such uncertainties in the context of DM-electron scattering. The differences in the kinematics of DM-nuclear scattering compared to the DM-electron scattering warrants a dedicated analysis to explore these effects. In the latter case, the momentum transfer between the DM and the electron is dominated by the electron's momentum. However, DM masses that sit at the experimental threshold are especially sensitive to the high-velocity tail of the halo's VDF. Currently, there are several experiments with dedicated searches for DM-electron scattering such as SENSEI~\cite{Tiffenberg:2017aac,Crisler:2018gci,Abramoff:2019dfb,Barak:2020fql}, SuperCDMS-HV~\cite{Agnese:2018col}, DAMIC~\cite{Aguilar-Arevalo:2019wdi}, Edelweiss~\cite{Arnaud:2020svb}, that have recently taken or are actively taking data. The next decade will see the arrival of larger silicon CCD experiments such as the 1-kg DAMIC-M~\cite{Lee:2020abc,Castello-Mor:2020jhd} and the proposed 10-kg Oscura~\cite{Oscura} experiments. Understanding the influence of astrophysical uncertainties on the observables of DM-electron scattering will allow us to interpret the results of these experiments in a more robust way.

This paper is organized as follows. We begin $\S$\ref{sec:DMeformalism} with an overview of the DM-electron scattering formalism, where we define the various ingredients that appear in the rate calculation. In $\S$\ref{sec:halomodels}, we introduce the halo models under study: the Standard Halo Model, the Tsallis model, and an empirical model derived from numerical simulations. We follow with a discussion of the astrophysical parameters in $\S$\ref{sec:astro_params}, where we briefly review how these parameters are measured and suggest new, updated values for the circular velocity $v_0$, escape velocity $\vesc$ and DM density $\rhoDM$. In $\S$\ref{sec:results}, we present the results of the effects of the underlying DM halo VDF on the observables of DM-electron scattering. We conclude in $\S$\ref{sec:conclusions}.

%%%%%%%%%%%%%%%%%%%%%%%%%%%%%%%%%%%%%%%%%%%%%
\section{Dark Matter - Electron Scattering Formalism}
\label{sec:DMeformalism}
%%%%%%%%%%%%%%%%%%%%%%%%%%%%%%%%%%%%%%%%%%%%%
In this section, we give a brief review of the formalism behind DM-electron scattering; more details can be found in~\cite{Essig:2015cda}. For a spherically-symmetric DM velocity distribution, the differential rate for DM-electron scattering in a crystal target is given by
\begin{align}
    \label{eq:diff_rate}
  \frac{dR}{ d\ln E_{e}  }  & =  N_{\rm cell}\frac{\rhoDM}{m_\chi}\frac{\overline\sigma_e \alpha m_e^2}{\mu_{\chi e}^2} \int d\ln q\, \frac{E_e}{q}\left[ |F_{\rm DM}(q)|^2
              |f^{\rm crystal}(E_e, q)|^2
                   \eta(v_{\rm min}(q, E_e)) \right] \,,
\end{align}
where $\rhoDM$ is the local DM density, $m_{\chi}$ is the DM mass, $q$ is the momentum transfer between the DM and the electron, $E_e$ is the energy transferred to the electron, $\alpha\simeq 1/137$ is the fine-structure constant, while $m_e$ and $\mu_{\chi e}$ are the mass of the electron and reduced mass of the DM-electron system, respectively. The material dependence of the rate is encoded in $f^{\rm crystal}(E_e, q)$, which is the crystal form-factor for an electronic transition of $E_e$ and $q$, and $N_{\rm cell}\equiv M_{\rm target}/M_{\rm cell}$, which counts the number of unit cells per mass in the crystal; for silicon, $M_{\rm cell}=2\times m_{\rm Si}=52.33$ GeV. We use the values of $f^{\rm crystal}(E_e, q)$ calculated from {\tt QEdark}~\cite{Essig:2015cda} for silicon.\footnote{We choose silicon for illustrative purposes. Nonetheless, our calculations can be performed for any semiconductor material with qualitatively the same conclusions. Similar calculations can also be made for noble liquid targets such as xenon and argon; however, noble liquids will be less sensitive to changes in the DM VDF compared with semiconductor crystals.}

The momentum-dependent DM-electron scattering cross section due to a mediator particle of mass $m_V$ is factorized as $\sigma_e(q)=\overline\sigma_e\times |F_{\rm DM}(q)|^2$, where the interaction strength is given by
\begin{equation}
    \overline\sigma_e\equiv\frac{\mu_{\chi e}^2\overline{|{\cal M}_{\rm free}(\alpha m_e)|^2}}{16\pi m_\chi^2 m_e^2}\, ,
\end{equation}
while the DM form factor
\begin{align}
  |F_{\rm DM}(q)|^2\equiv
  \frac{ \overline{ |\mathcal{M}_{\rm free}(q)|^2} } { \overline{ |\mathcal{M}_{\rm free}(\alpha m_e)|^2}} =
  \begin{cases}
    1 &,~m_V\gg \alpha m_e~(\text{contact})\\
    \left(\frac{\alpha^2m_e^2}{q^2}\right)^2 &,~m_V\ll \alpha m_e~(\text{light med.})  \end{cases}\,
\end{align}
expresses the momentum dependence of the interaction -- the top line corresponds to a contact interaction via a heavy vector mediator, while the bottom line corresponds to a long-range interaction via the exchange of a massless or ultralight vector mediator.
${\cal M}_{\rm free}(\vec q)$ is the matrix element for the free-elastic scattering of DM with an electron, while $\overline{|{\cal M}|^2}$ denotes the absolute square of $\cal M$ averaged over initial and summed over final particle spins.
The DM halo velocity profile is encoded in the velocity average of the inverse speed, defined as
\begin{align}
    \eta(\vmin) = \int \frac{d^3v}{v} f_\chi(v) \: \Theta(v - \vmin)
    \label{eq:eta_int}
\end{align}
where $\Theta$ is the Heaviside step function. Energy conservation is captured by the minimum speed needed by the DM particle in order for an electron to gain energy $E_e$ with momentum transfer $q$,
\begin{equation}
\vmin(E_e,q)=\frac{E_e}{q}+\frac{q}{2m_\chi},\
\end{equation}
while $f_\chi(v)$ is the velocity distribution of dark matter in our Galaxy and is the focus of our work. Note that $f_\chi(v)$ is evaluated in the detector (Earth) frame and is sensitive to the Earth's velocity. This dependence manifests itself as an annual modulation of the count rate, which we will discuss in more detail in $\S$\ref{sub:earth}.

DM-electron scattering experiments measure electron-hole pairs $N_e$, rather than electron energy $E_e$. The conversion of $E_e$ to $N_e$ involves secondary scattering processes, the exact modeling of which is ongoing and beyond the scope of this work. Instead, we use a linear response model given by
\begin{equation}
    N_e=1+\left\lfloor\frac{E_e-E_g}{\varepsilon}\right\rfloor\, ,
\end{equation}
that provides a reasonable estimate for the conversion of $E_e$ to $N_e$.
Here, $E_g$ is the band-gap energy of the detector material and $\varepsilon$ is the energy needed to produce an additional electron-hole pair above the band-gap. The floor operator $\lfloor x \rfloor$ rounds $x$ down to the nearest integer. The corresponding values for silicon are measured at $E_g\simeq 1.2$ eV and $\varepsilon\simeq3.8$ eV~\cite{Vavilov_1962,Rodrigues:2020xpt}.

We end this section with a discussion about the kinematics of DM-electron scattering and its dependence on the DM velocity. In DM-electron scattering the target is a bound electron; therefore, the electron does not have a definite momentum but rather can take on a wide range of values. One can relate the momentum transfer $q$ to $E_e$ via energy conservation, according to
\begin{equation}
    E_e=\vec q\cdot \vec v-\frac{q^2}{2m_\chi}\, .
    \label{eq:Ee}
\end{equation}
Since the velocity of a bound electron is $v_e\sim Z_{\rm eff}\alpha \gg v_\chi\sim 10^{-3}$, the typical momentum transfer is set by the electron's momentum. Therefore, $q_{\rm typ}\sim Z_{\rm eff}\alpha m_e$, where $Z_{\rm eff}=1$ for the outer shell electrons and increases for inner shells.

The energy transfer in eq.~\ref{eq:Ee} is dominated by the first term on the right-hand side for $m_\chi$ well-above the threshold energy, $E_{\rm min}=\frac{1}{2}\mu_{\chi N}^2 v^2\simeq \frac{1}{2}~{\rm eV}\left(\frac{m_\chi}{{\rm MeV}}\right)$, where $\mu_{\chi N}$ corresponds to the reduced mass of the DM-nucleus system. In this regime, the minimum momentum transfer required to obtain an energy transfer $E_e$ is given by $q_{\rm min}=E_e/v\sim E_e/(Z_{\rm eff}\cdot 4{~\rm eV)}\times q_{\rm typ}$. What we learn from this scaling is that the typical momentum transfer leads to transitions of a few eV; a more energetic transition requires higher DM velocity, or, alternatively, a high electron momentum which is suppressed. From this, we can deduce that the rate is sensitive to the DM halo velocity profile and in what follows we quantify this sensitivity.
%%%%%%%%%%%%%%%%%%%%%%%%%%%%%%%%
\section{Halo Models}
\label{sec:halomodels}
%%%%%%%%%%%%%%%%%%%%%%%%%%%%%%%%
In this section, we describe the halo models under study: the Standard Halo Model, the Tsallis model, and an empirical model derived from numerical simulations. We present the functional forms for the corresponding halo distributions and motivate the choice of each model. Note that although we present the functional forms of the VDFs in the galactic rest frame, the quantity that enters the rate calculation must be evaluated in the Earth's rest frame. The normalization of each halo model satisfies $\int f(v) d^{3}v =1$. We show the general form for $\eta$ and $v^2 f(v)$ for the halo models under discussion in figs.~\ref{fig:eta_panel} and~\ref{fig:v2f_panel} along with their dependence on astrophysical parameters, which we will discuss in more detail in $\S$~\ref{sec:astro_params}.

\vskip 0.3 cm

%%%%%%%%%%%
\subsection{Standard Halo Model}
\label{sec:SHM}
The Standard Halo Model (SHM) \cite{drukier1986detecting} has been featured extensively in several analyses of DM detection experiments, serving as a convenient option for the modeling of the dark halo. The model assumes an isothermal spherical distribution for the dark matter in the Galaxy with a density profile that scales with the distance from the center of the Galaxy as $r^{-2}$, to match the behavior borne out in the observations of galactic rotation curves. Under these assumptions, an isotropic Maxwell-Boltzmann (MB) velocity distribution emerges self-consistently as the solution to the collisionless steady-state Boltzmann equation \cite{lisanti2017lectures}.
A physical cutoff for the distribution is imposed at the local escape speed $\vesc$ while, within this approximation of a singular isothermal sphere, its most probable speed $v_{0}$ is identified with the local circular speed~\cite{kuhlen2010dark}.
The MB distribution is given by
\begin{align}
  f_{\rm MB}(\vec{v}) &\propto \begin{cases}
  e^{-|\vec v|^2/v_0^2} &|\vec{v}| < \vesc \\
  0 &|\vec{v}| \geq \vesc.
  \end{cases}
\end{align}
When boosted to the detector's (Earth) frame, the distribution takes the form
%%%%%%%%%%%
\begin{align}
f_{\rm SHM}(\vec v) =
\frac{1}{K} e^{-|\vec{v}+\vec v_{E}|^2/v_0^2}\Theta(\vesc - |\vec v+\vec v_E|),
\label{eq:SHM}
\end{align}
where $v_E$ is the Earth's Galactic velocity and
\begin{align}
K_{\rm SHM} = v_0^3 \left(\pi^{3/2} \erf(\vesc/v_0) - 2 \pi \frac{\vesc}{v_0} e^{-\vesc^2/v_0^2}\right)
\end{align}
is the normalization constant that results from enforcing that $\int f(v) d^{3}v =1$.

For a spherically-symmetric system with an isotropic velocity tensor, the Eddington formula \cite{eddington1916distribution} can be used to derive the DM phase-space distribution function from first principles \cite{binney2008galactic}. This method warrants that the inferred speed distribution flows naturally from the choice of the spatial density profile of the DM component, providing a self-consistent connection to the total gravitational potential of the Galaxy. \footnote{For a review on the Eddington approach and its anisotropic extensions see Ref.\cite{lacroix2018anatomy} and references therein.}
Though for the scope of this work we are solely interested in distributions for which there is no preferred direction for the velocities of the DM particles at the solar vicinity,
if we choose to retire the assumption of isotropy we come across anisotropic realizations of the DM VDF. Such models break the 1-1 direct relation between the density distribution and the velocity distribution of the dark matter \cite{green2017astrophysical} and can have important implications for the DM detection observables \cite{ullio2001velocity, bozorgnia2013anisotropic, fornasa2014self, evans2019refinement}. For some synthetic halos that have been found to exhibit such anisotropic behaviors at a varying degree, see representative examples in refs.~\cite{vogelsberger2009phase, ling2010dark, kuhlen2010dark, sparre2012behaviour,bozorgnia2017implications}.

The average inverse speed for the SHM can be calculated analytically in two kinematic regimes,
\begin{enumerate}
\item $v_{\rm min}<v_{\rm esc}-v_E$,
\item $ v_{\rm esc}-v_E< v_{\rm min}<v_{\rm esc}+v_E$
\end{enumerate}

\noindent where $\vesc,~v_E,~v_{min}>0.$

\noindent Following  ref.~\cite{Lewin:1995rx}, we arrive at
\begin{eqnarray}
 \eta_1(v_{\rm min})&=&\frac{v_0^2\pi}{2 v_E K}\left(-4e^{-v_{\rm esc}^2/v_0^2}v_E+\sqrt{\pi}v_0\left[\textrm{Erf}\left(\frac{v_{\rm min}+v_E}{v_0}\right)-\textrm{Erf}\left(\frac{v_{\rm min}-v_E}{v_0}\right)\right]\right)\\
  \eta_2(v_{\rm min})&=&\frac{v_0^2\pi}{2 v_E K }\left(-2e^{-v_{\rm esc}^2/v_0^2}(v_{\rm esc}-v_{\rm min}+v_E)+\sqrt{\pi}v_0\left[\textrm{Erf}\left(\frac{v_{\rm esc}}{v_0}\right)-\textrm{Erf}\left(\frac{v_{\rm min}-v_E}{v_0}\right)\right]\right)\nonumber
\end{eqnarray}
where the subscript corresponds to the case number. Note that the two cases converge to the same value for $v_{\rm min}=v_{\rm esc}-v_E$.

The SHM does not reproduce the features of the VDFs inferred by DM-only simulations, in particular when it comes to the high-velocity tail of the distribution~\cite{fairbairn2009spin, vogelsberger2009phase, march2009inelastic, kuhlen2010dark, mao2013halo}, although it is in better alignment with VDFs resulting from simulations that include baryonic physics \cite{ling2010dark, pillepich2014distribution, bozorgnia2016simulated, butsky2016nihao, kelso2016impact, sloane2016assessing} (see detailed discussion in  $\S$\ref{sub:sims}). In general, it is safe to say that the SHM does not provide a realistic description of the DM distribution in the Galactic halo, thus motivating the discussion of alternative models for the DM VDF in the following subsections.

%%%%%%%%%%%
\subsection{Tsallis Model}
\label{sec:tsa}
%%%%%%%%%%%
The use of the Tsallis VDF~\cite{Tsallis1988} was proposed to describe data from numerical $N$-body simulations with baryons \cite{Hansen_2006,Vergados:2007nc,Ling:2009eh}. While the SHM follows the assumption of standard Boltzmann-Gibbs entropy, $S_{BG} = -k \sum p_i \ln p_i$, where $p_i$ is the probability for a particle to be in state $i$, the Tsallis model results from Tsallis statistics \cite{Tsallis1988}, a generalization of the standard definition of entropy via the introduction of the entropic index \textit{q},
\begin{align}
  S_q &= -k \sum_i p_i^q \ln_q p_i,
\end{align}
where $\ln_q p = (p^{1-q}-1)/(1-q)$ is defined as the $q$-logarithm \cite{hansen2005dark}. By taking the limit $q \to 1$ we recover the standard Boltzmann-Gibbs entropy. The Tsallis VDF is the exact analog of the SHM but in the generalized Tsallis regime, and describes non-extensive systems. With this Tsallis entropy, one arrives at the Tsallis VDF,
\begin{align}
  f_{\rm Tsa}(\vec{v}) &\propto \begin{cases}
  \left[1-(1-q)\frac{\vec{v}^2}{v_0^2}\right]^{1/(1-q)} &|\vec{v}| < \vesc \\
  0 &|\vec{v}| \geq \vesc.
  \end{cases}
\end{align}

An argument in favor of using the Tsallis model to describe the velocity profile of the DM halo is due to the fact that this distribution has the escape speed already built-in for the $q<1$ regime, thus the cutoff does not have to be artificially added as in the SHM case. For $q<1$, $\vesc$ is fixed by $q$ and $v_0$ through the relation $\vesc^2=v_0^2/(1-q)$, but for $q>1$ the escape speed is an independent parameter and still has to be added manually. By examining the velocity distributions inferred by simulations~\cite{ling2010dark}, we observe that the high-speed tails exhibit a steeper fall-off and a smoother transition to zero compared to the abrupt fall-off of the SHM VDF. This is a feature that has been associated with relaxed, collisionless structures and is a key characteristic of the Tsallis VDF~\cite{lima2001nonextensive,hansen2005dark, hansen2006universal}.

The standard parameter choices of $v_0 = 220$ km/s and $\vesc = 544$ km/s correspond to $q =0.836$, and our suggestions for $v_0 = 228.6$ km/s and $\vesc = 528$ km/s lead to $q = 0.813$. Ref. \cite{Ling:2009eh} fits the Tsallis model to a simulation using the RAMSES code~\cite{teyssier2002} with $R_0=8.0$ kpc and yields $q=0.773$ and $v_0=267.2$ km/s, corresponding to $\vesc = 560.8$ km/s; we note that these values are well within the error margins for the standard parameters.

%%%%%%%%%%%
\subsection{Empirical Models}
\label{sec:msw}
%%%%%%%%%%%
\label{sub:sims}
Cosmological simulations have been instrumental in shaping our understanding of physical processes that one cannot directly observe, such as the non-linear evolution and formation of cosmological structures. A detailed overview of the relevant developments in numerical cosmological simulations can be found in ref.~\cite{vogelsberger2020cosmological}. In this section, we briefly review how cosmological simulations are constructed and their applications to deriving the DM VDF.

Cosmological simulations are broadly divided into two categories: $N$-body methods, where simulation codes are employed to describe the dynamics of pure collisionless dark matter, and the full hydrodynamical approach where numerical techniques are implemented to study the behavior of the DM component in the presence of baryons. Both classes of simulations trace the phase-space density evolution of the DM fluid, facilitating the extraction of a local VDF associated with each cosmological setup.
As revealed by several studies \cite{fairbairn2009spin, vogelsberger2009phase, march2009inelastic, kuhlen2010dark, mao2013halo} that used data from DM-only simulations, the general trend followed by the associated local VDFs translates into a considerable deviation from the SHM, with the simulation distributions exhibiting a suppressed peak compared to the shape of the truncated Maxwellian and a surplus of particles at the high-speed tail.

Although a layer of appreciable complexity is added to the simulations when incorporating baryonic processes, this step is crucial for replicating realistic conditions in the simulated cosmological environments, and indispensable in order to make robust predictions.
Ref.~\cite{bozorgnia2017implications} provides a comprehensive review of several analyses \cite{ling2010dark, pillepich2014distribution, bozorgnia2016simulated, butsky2016nihao, kelso2016impact, sloane2016assessing} that drew data from various hydrodynamical simulations. The different design of these simulations can lead to large variations in their
final products, including discrepancies in the respective local speed distributions inferred from each simulation, as well as a halo-to-halo scatter concerning the local speed distributions within the same simulation suite. Despite their individual differences, the simulations that include baryonic effects shift the peak of the velocity distributions to higher speeds and, in most cases, show a better convergence with the standard Maxwellian distribution, compared to their DM-only counterparts.
Nevertheless, an exact shape for the DM VDF has yet to be captured, and more higher-resolution simulations with an accurate implementation of baryonic physics are required for a universal agreement to be established.

There are several empirical approaches to the DM halo velocity profile derived from various cosmological simulations~\cite{Lisanti:2010qx,Sloane:2016kyi,Hryczuk:2020trm}. For our purposes, we will be focusing on the empirical model of refs.~\cite{Mao:2012hf, Mao:2013nda} based on the DM-only simulation packages Rhapsody \cite{wu2013rhapsody} and Bolshoi \cite{klypin2011dark}. Previous work has investigated the sensitivity of DM-electron scattering on the DM velocity profile using empirical models derived from the IllustrisTNG simulations~\cite{Hryczuk:2020trm}. Our work is complementary given that we consider a different set of halo-models. In the Galactic rest frame, the empirical halo velocity distribution is given by

\begin{equation}
    f_{\rm emp}(\vec v)\propto
    \begin{cases}
    e^{-|\vec v|/v_0}\left(v_{\rm esc}^2-|\vec v|^2\right)^p,  &|\vec{v}| < \vesc \\
    0, &|\vec{v}| \geq \vesc ,
    \label{eq_Mao}
    \end{cases}
\end{equation}
where the circular speed $v_{0}$ and the power $p$ are allowed to vary.

This function is composed out of two terms: an exponential term, that accounts for the halo's velocity anisotropy, and a cutoff term that regulates the number of particles that are bound to the Galactic potential.
After fitting each simulated halo with the Navarro-Frenk-White (NFW) density profile, the authors argue that the dominant theoretical uncertainty on the VDF ---and correspondingly the scattering event rate--- can be traced to the term $r/r_{s}$ in the NFW profile, {\it i.e.} the ratio of the radial position where the VDF is measured to the scale radius of the halos' density profile. This quantity points directly to the VDF's connection to the gravitational potential, and ultimately determines the shape of the VDF that exhibits a broader peak and steeper tail, in contrast to the SHM.

Ref.~\cite{pillepich2014distribution} studies the effects of baryonic feedback on the local DM VDF, comparing the resulting distributions extracted from halos in the Eris~\cite{guedes2011forming} and ErisDark simulations.
Both are well fitted by the empirical distribution in eq.~\ref{eq_Mao}, while the best-fit parameters are ($v_0, p$) = (330 km/s, 2.7) for Eris, and (100 km/sec, 1.5) for its DM-only counterpart. We use $p=1.5$ for our fiducial model.

\vskip 0.3 cm

%%%%%%%%%%%
\subsection{Debris Flows from Observational Data}
\label{sec:df}
%%%%%%%%%%%
Observational missions allow us to use stars to trace the evolution and distribution of DM structures in the Milky Way halo.
In ref.~\cite{herzog2018empirical} the authors suggest that an empirical velocity distribution for the DM can be determined through the study of metal-poor stars in the solar vicinity. After testing the correlation between the stellar and the DM distributions using the Eris hydrodynamic simulation~\cite{guedes2011forming}, they extrapolated this correspondence to infer the VDF for the virialized DM component using SDSS data~ \cite{carollo2010structure}. The resulting distribution exhibits differences when contrasted with the SHM, with notably the anisotropic behavior of the metal-poor stellar distribution.
The correlation between the stellar and DM kinematic distributions demonstrated in~\cite{herzog2018empirical} prompted the reconstruction of the local velocity distribution associated with the smooth DM component in~\cite{herzog2018metal}, with the use of metal-poor stellar populations from the RAVE~\cite{kunder2017radial}-TGAS~\cite{michalik2015tycho} catalog.

But no VDF can be complete unless it accounts for the features associated with DM substructure.
The study of a combined SDSS~\cite{ahn2012ninth}-{\it Gaia} DR2~\cite{brown2018gaia} sample of accreted stars in \cite{necib2019inferred} indicates that a non-negligible fraction of the DM in the halo could be comprised of debris flow --- provided that the selected stars adequately trace the DM component.
Once this anisotropic stellar population is included in the analysis, the resulting velocity distribution deviates from the SHM prediction.
Earlier works \cite{lisanti2012dark, kuhlen2012direct} have studied the formation and emergence of this velocity phase-space substructure in $N$-body Via Lactea II (VL2) simulations~\cite{diemand2008clumps, kuhlen2008via}, where the debris flow was identified as spatially-homogeneous material that accumulated from tidally-stripped luminous satellites that fell into our Galaxy. We used the results from refs.~\cite{necib2019inferred,github} to simulate the debris flow component of the DM VDF.
We find that the addition of this substructure component has a subdominant effect in DM-electron scattering compared to varying the astrophysical parameters of the halo model. The reason for this is twofold: the debris flow is a sub-component of the total distribution, and it contributes mostly at low velocities. Therefore, we do not discuss the effect of debris flows in detail. Nevertheless, the sensitivity of DM-electron scattering to the DM velocity distribution can be exploited to provide information about the underlying DM substructure; the effects of different types of DM substructure, including debris flows and stellar streams, are explored in the context of DM-electron scattering in \cite{buch2020dark}.

\section{Astrophysical Parameters}
\label{sec:astro_params}
%%%%%%%%%%%%%%%%%%%%%%
\begin{figure}[h]
    \centering
    \includegraphics[width=0.48\textwidth]{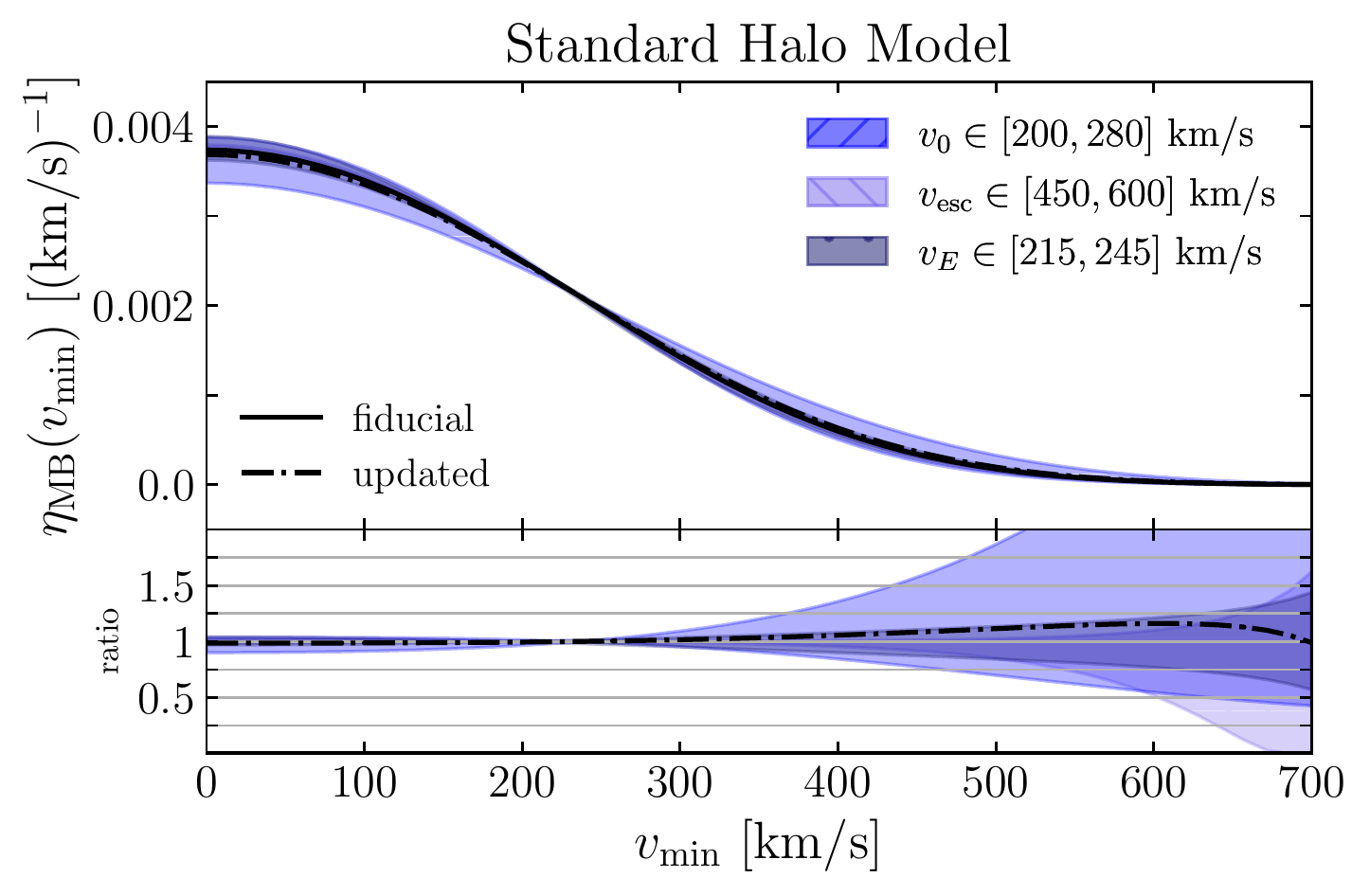}
    \includegraphics[width=0.48\textwidth]{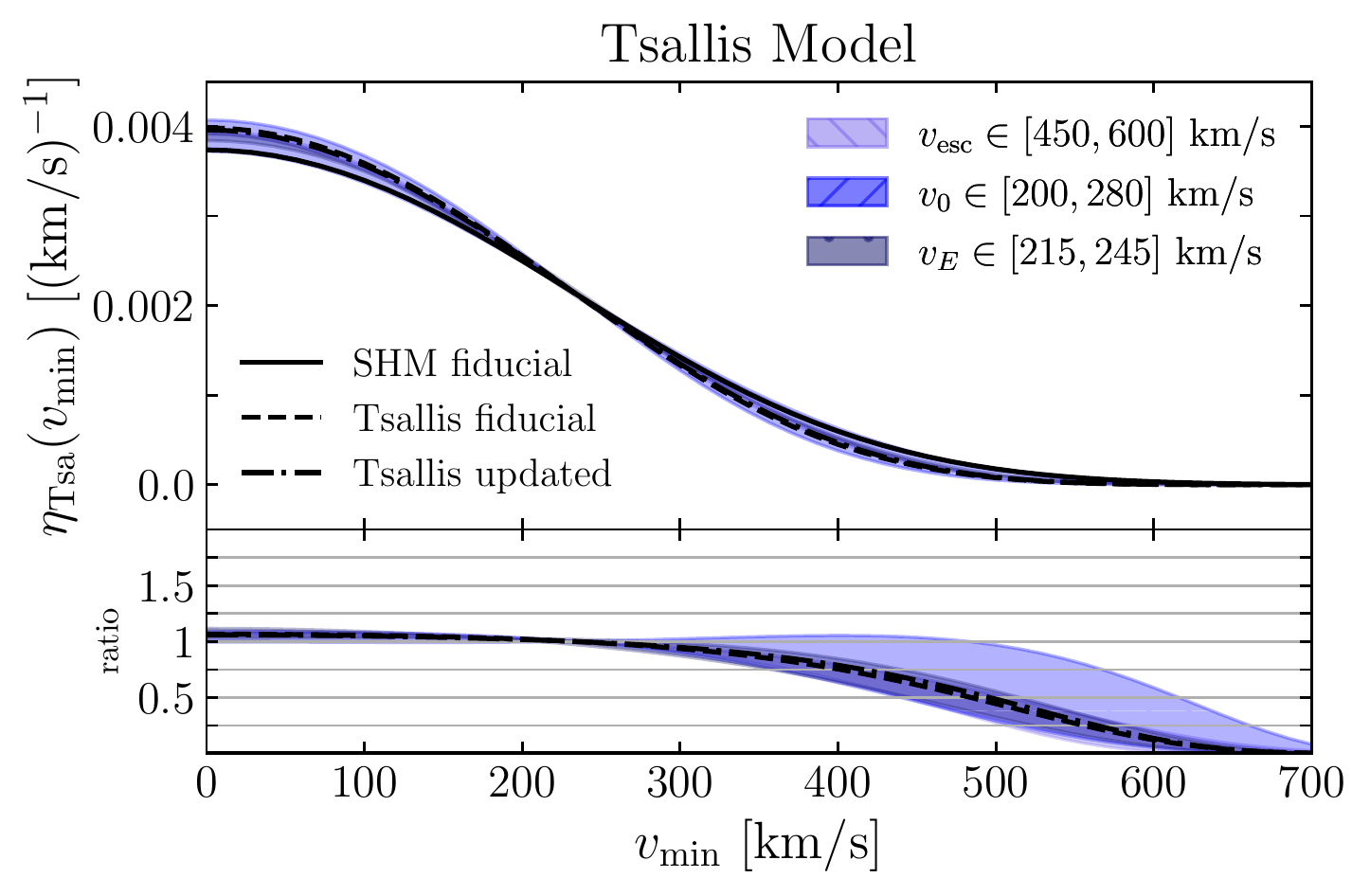}
    \includegraphics[width=0.48\textwidth]{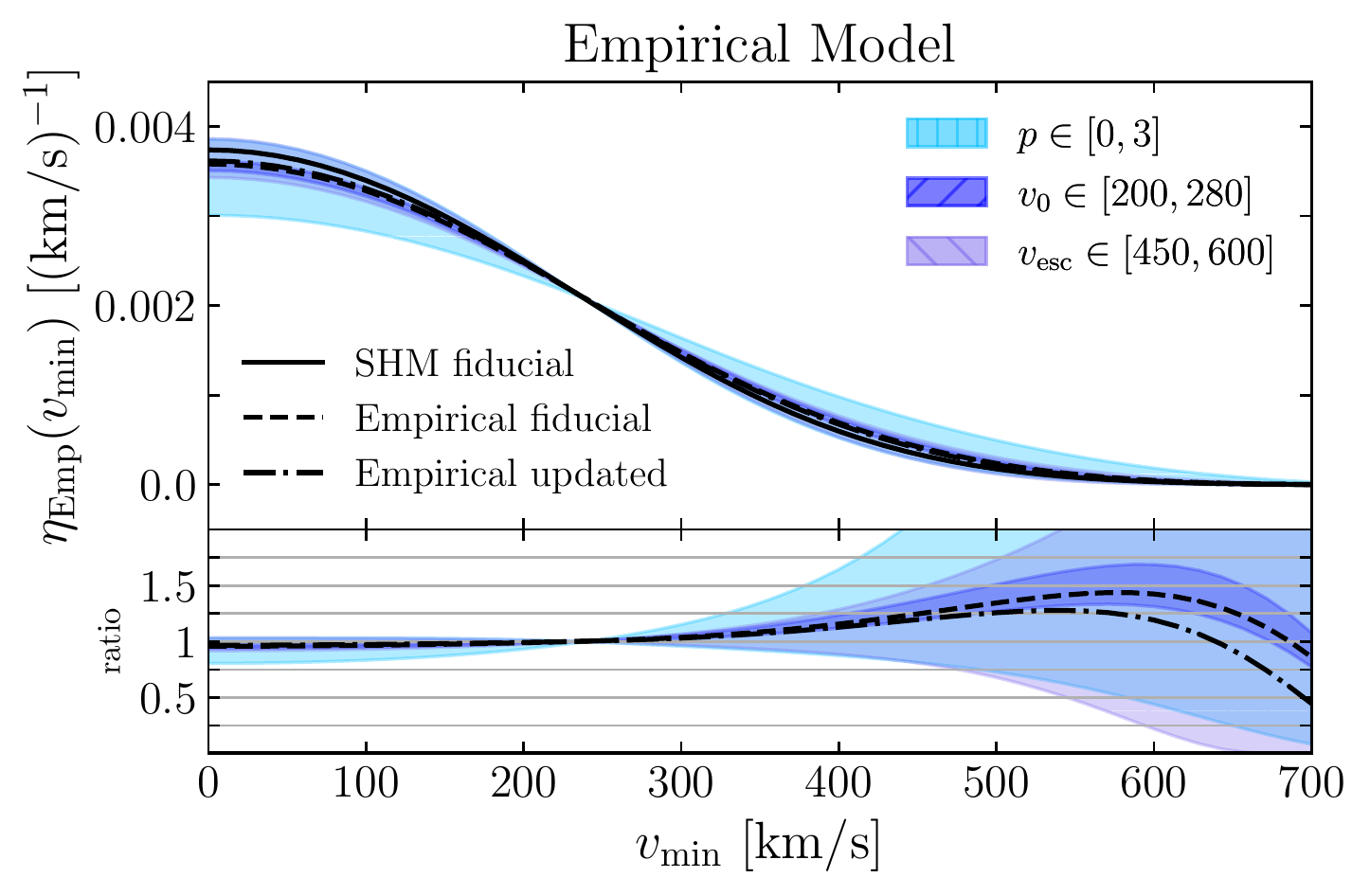}
    \includegraphics[width=0.48\textwidth]{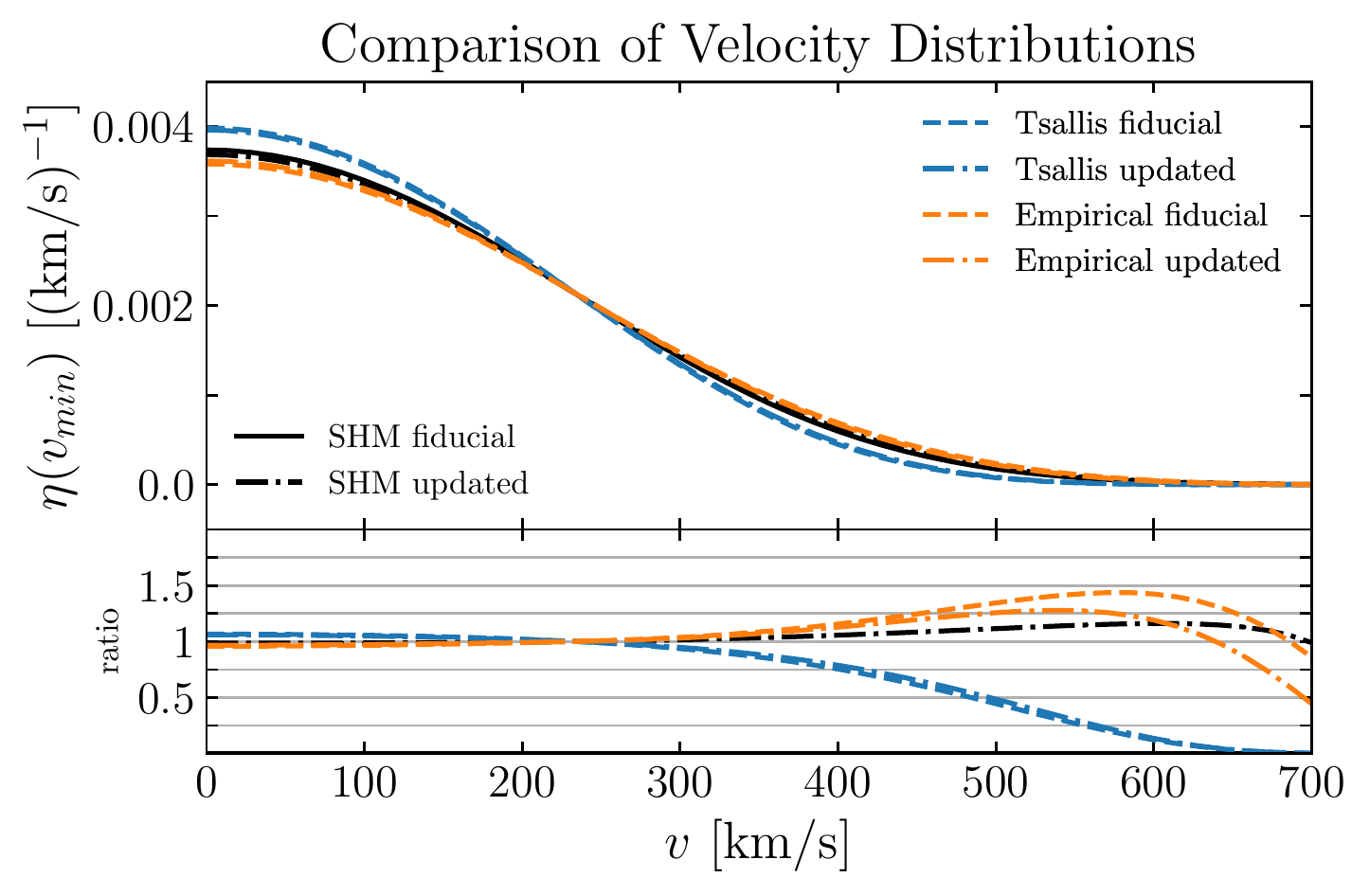}
    \caption{Integrated $\eta(v_{\rm min})$, as defined in eq.~\ref{eq:eta_int}, for the three velocity distributions varied over their parameters. In the bottom, right panel, we show a comparison between the three models at their fiducial values. The black solid line corresponds to the SHM, while the dashed lines represent the Tsallis and empirical distributions, all evaluated at their fiducial parameters. The dash-dotted curves in each panel result from evaluating each model at their updated parameters. When varying over one parameter, the remaining parameters are kept fixed to their fiducial values.}
    \label{fig:eta_panel}
\end{figure}

\begin{figure}[h]
    \centering
    \includegraphics[width=0.48\textwidth]{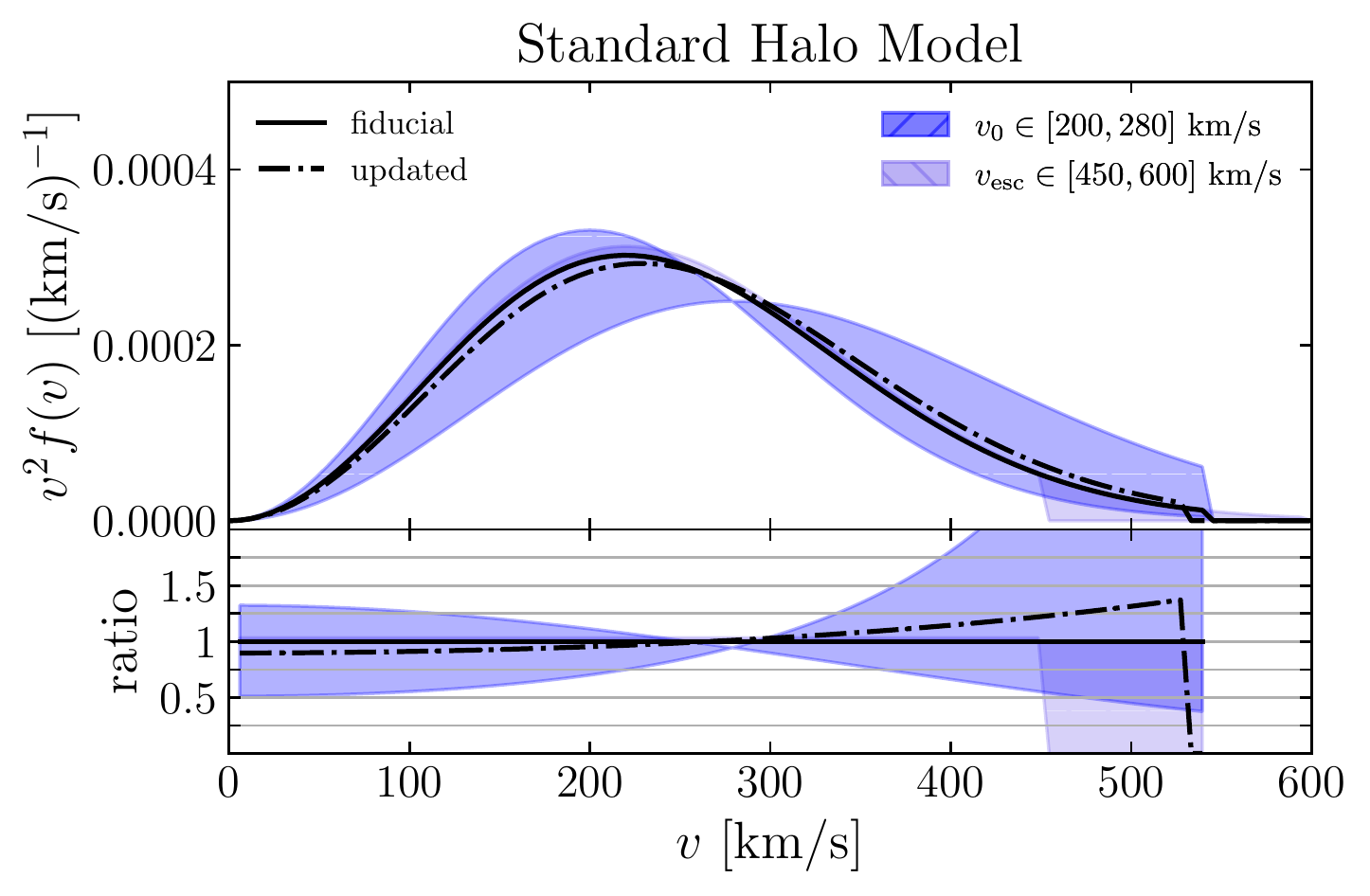}
    \includegraphics[width=0.48\textwidth]{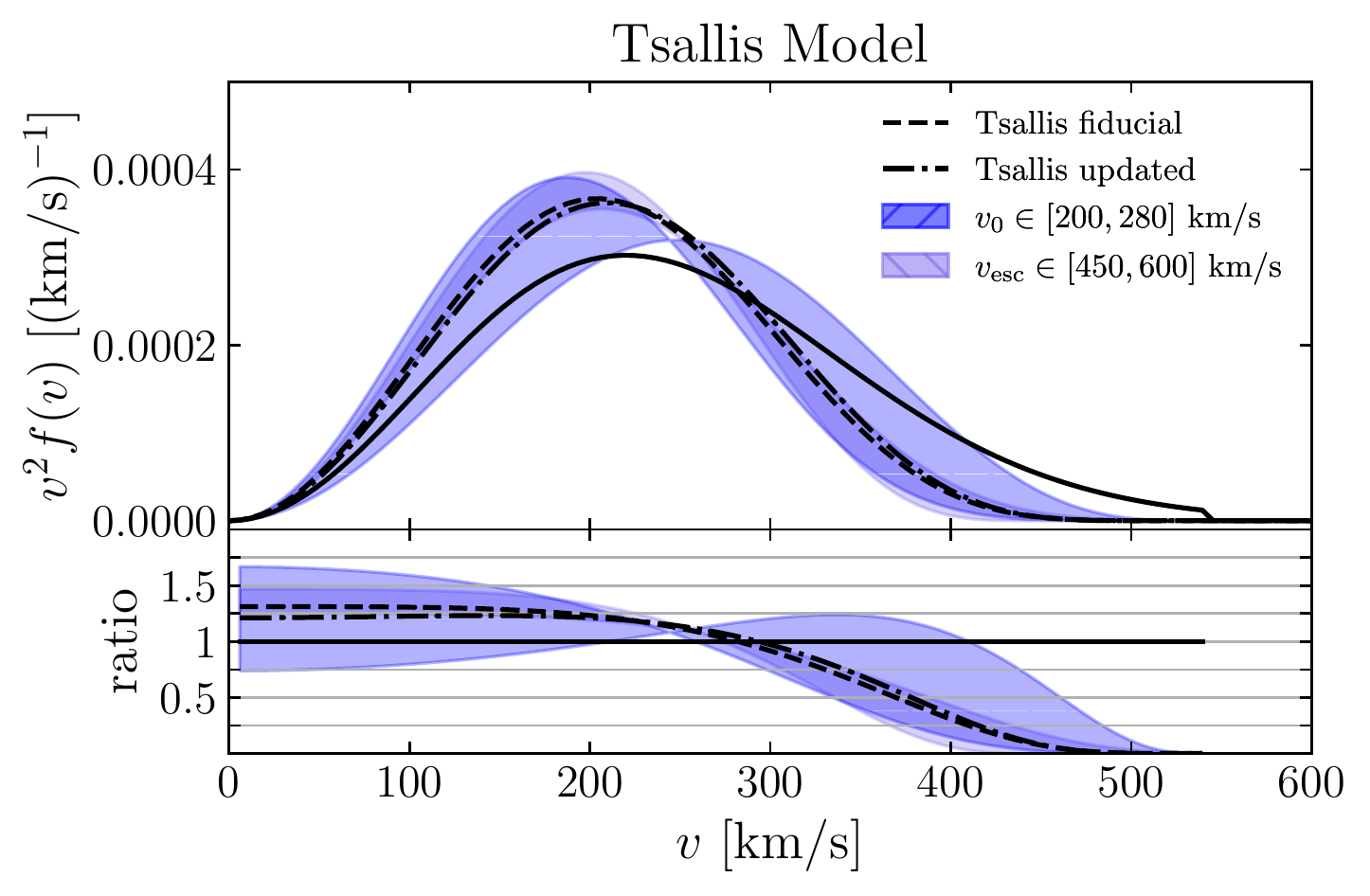}
    \includegraphics[width=0.48\textwidth]{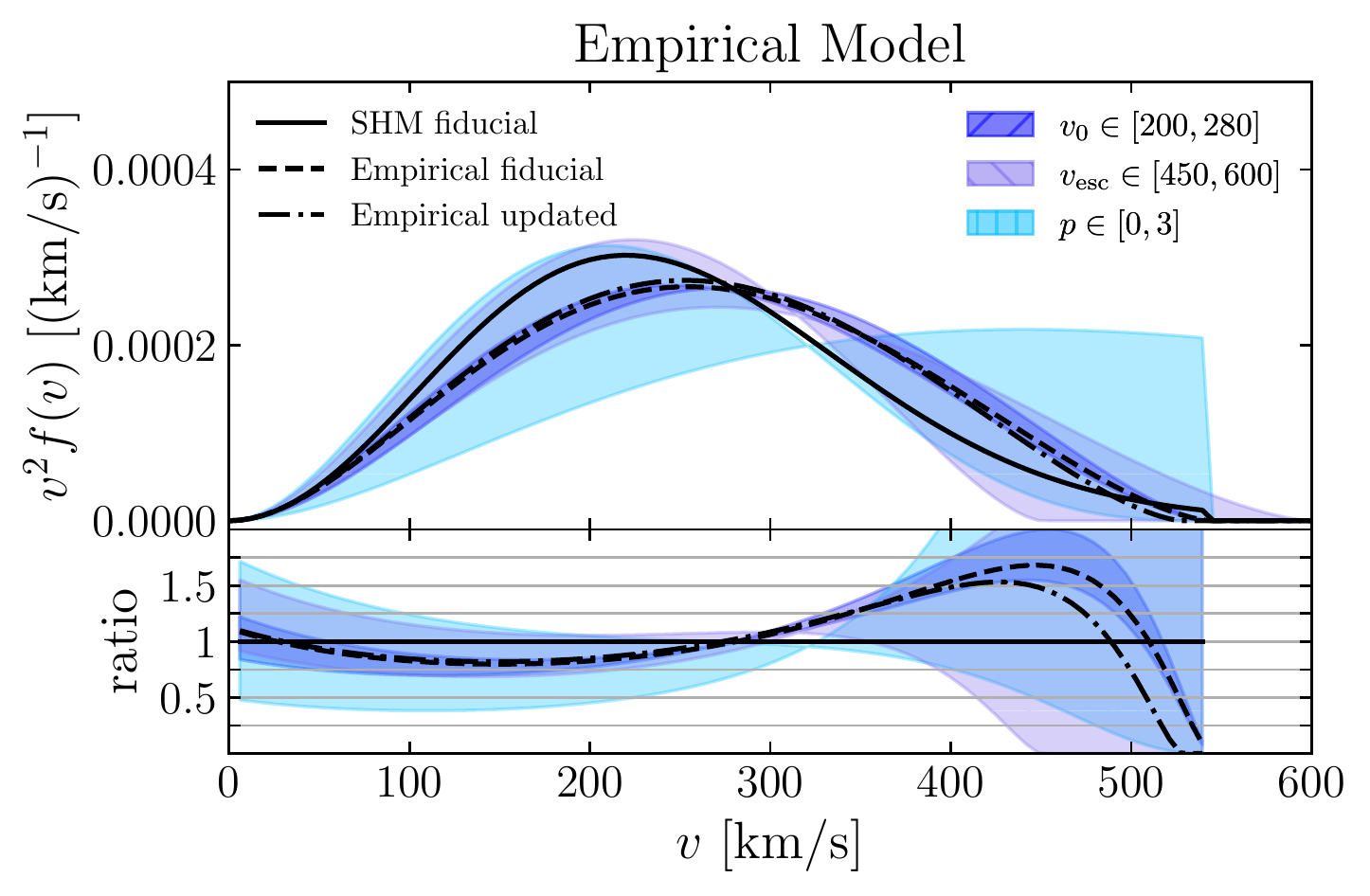}
    \includegraphics[width=0.48\textwidth]{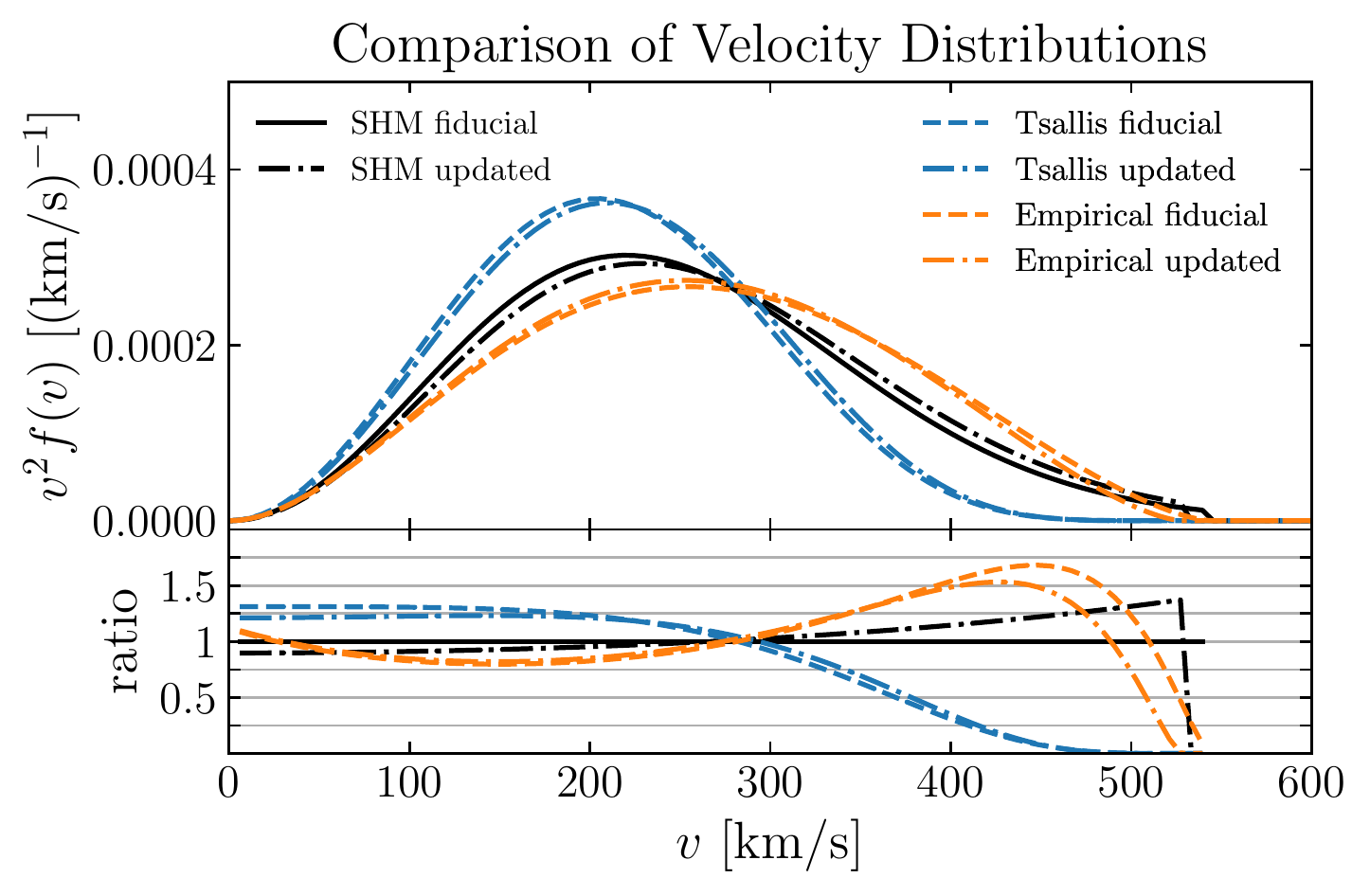}
    \caption{ $v^2f(v)$ for the three velocity distributions varied over their parameters. The black solid line corresponds to the SHM, while the dashed lines represent the Tsallis and empirical distributions, all evaluated at their fiducial parameters. The dash-dotted curves in each panel result from evaluating each model at their updated parameters. The panel on the bottom right corner shows the comparison among the three models. When varying over one parameter, the remaining parameters are set to their fiducial values.}
    \label{fig:v2f_panel}
\end{figure}
The main astrophysical quantities involved in the calculation of the DM-electron scattering recoil spectrum are the local DM density $\rhoDM$, the circular velocity $v_{0}$, the escape velocity $\vesc$, and the Earth velocity $v_{E}$. Here we review how these parameters are measured, and summarize the corresponding sets of measurements provided by currently ongoing experiments. We recommend the use of parameter values from these updated measurements.

%%%
\subsection{Local Dark Matter Density}
\label{sub:DM_density}
%%%
The local DM density, $\rhoDM$, appears as an overall scaling for the DM scattering rate. The value of $\rhoDM$ can be measured via both local \cite{kapteyn1922first, oort1960note, bahcall1984self, kuijken1991galactic, garbari2012new, bovy2012local, zhang2013gravitational} and global \cite{dehnen1998mass, weber2010determination, catena2010novel, mcmillan2011mass} methods that offer complementary information about the galactic structure, albeit employing independent techniques (see ref.~\cite{read2014local} for a thorough review on the subject, as well as an extensive list of the relevant measurements for the $\rhoDM$ parameter).

Local measurements utilize the vertical ---in the direction perpendicular to the plane of the Milky Way disk---  motions of stellar tracers, located within a small volume surrounding the solar circle, to
probe the gravitational potential of the Galaxy. To isolate the contribution of the DM component to the total Galactic potential, substantial modeling is required to assess how baryonic matter is distributed within the Galactic disk. The latter undertaking poses one of the greatest challenges in determining the local DM density in this way.
Considering that distinct populations of stars can result in different estimates for the local density, caution should be exercised also when choosing a representative stellar sample \cite{read2014local, buch2019using}.

Global measurements rely on a set of dynamical observables to fit parametrized models of the Milky Way. Specifying the model parameters in this fashion, among which the local DM density, posits spherical symmetry for the shape of the Galactic halo, notwithstanding potential departures from sphericity have been shown to give rise to systematic errors in the determination of $\rhoDM$ \cite{zemp2009graininess, pato2010systematic, bernal2014systematic}.

Local-type probes have been gaining ground lately with the advent of high-precision astrometric measurements spearheaded by the {\it Gaia} satellite.
Recently, refs.~\cite{buch2019using, widmark2019measuring} analyzed stellar data from the {\it Gaia} DR2 catalog and reported results over the interval (0.4 - 1.5) $\rm{GeV/cm^3}$, while according to ref.~\cite{Zyla:2020zbs} studies relying on global methods reported values that vary in the range (0.2 - 0.6) $\rm{GeV/cm^3}$.
Predating {\it Gaia's} second data release, ref.~\cite{sivertsson2018localdark} analyzed SDSS-SEGUE G-dwarf stellar data and reported the value of $0.46 ^{+0.07}_{-0.09} \rm{GeV/cm^{3}}$ for $\rhoDM$. This is the value we propose to the experimental collaborations for replacing the value of $\rhoDM = 0.4 \; \rm{GeV/cm^{3}}$, that has been the standard suggestion by the PDG~\cite{Tanabashi:2018oca} and has been in widespread use so far.

\subsection{Circular Velocity}
\label{sub:circ_v}

The rotation curve of the Milky Way $v_{c}(R)$ specifies how the orbital velocities of the stars and gas vary as a function of distance $R$ from the Galactic center, and consequently reflects how matter is distributed in the Galaxy. For an axisymmetric gravitational potential $\Phi$, the rotation curve is defined as $v_{c}(R) \equiv R^{2} \frac{\partial \Phi}{\partial R} \lvert_{z=0}$, where $z$ corresponds to the height above the Galactic disk plane. We define the circular velocity at the solar Galactocentric distance $R_{0}$ as $v_{0} \equiv v_{c}(R_{0})$.

There are two broad categories of methods devoted to the determination of the local circular velocity:  methods that are independent of $R_0$ and methods that rely on measurements of $R_0$. The first category sidesteps the complications related to the measurement of the Galactocentric radius and
entails direct probes of the local radial force. Examples of this method involve measurements of the GD-1 stellar stream that yield
$v_{0} = 221_{-20}^{+16}$~\cite{koposov2010constraining}, as well as the determination of the Oort constants from the analysis of {\it Hipparchos} data in~\cite{feast1997galactic} that lead to a value consistent with the established IAU recommendation of 220 km/s \cite{kerr1986review}.
Data taken from the Apache Point Observatory Galactic Evolution Experiment (APOGEE) yields a local circular velocity at $v_{0} = 218\pm{6}$~\cite{bovy2012milky}, without any assumptions about the relation between $v_{0}$ and the solar motion. However, this method is limited by the accuracy of the respective data sets and can be improved with increased statistics.

The second category relies on measurements of $R_0$ and, in principle, inherits the associated uncertainties. However, $R_0$ is increasingly well-measured
(see {\it e.g.}~\cite{bland2016galaxy} and references therein), and therefore the values of $v_0$ derived from this method have progressively smaller uncertainties.
 Given a fixed value for $R_0$, one can then infer a value for $v_0$.
For example, one can measure the Sun's motion with respect to a fixed reference point, relative to the Milky Way's center.
The most prominent point of reference to date is Sagittarius $A^{\star}$, the massive black hole radio source at the center of the Galaxy. The apparent motion of Sagittarius $A^{\star}$ in the galactic plane is primarily influenced by the effects of the solar orbit around the Galactic center~\cite{reid2004proper}. This observation can be translated into a constraint for the angular motion of the Sun, $\Omega_{0} = (v_{0} + V_{\odot})/R_{0}$, where $V_{\odot}$ corresponds to the component of the solar peculiar motion with respect to the Local Standard of Rest (LSR) in the direction of Galactic rotation, a quantity whose measurement introduces another source of uncertainty.\footnote{The $V_{\odot}$ components are challenging to determine due to the dependence of their associated asymmetric drift on the velocity dispersion of the stellar sample invoked to measure the solar peculiar motion vector ${\vec{v}_\odot}$~\cite{stromberg1946motions,
schonrich2010local}.
The authors in~\cite{ schonrich2010local}, by studying stellar data from the {\it Hipparchos} catalogue, inferred the stellar velocity distributions based on a model that takes into account the Galaxy's chemodynamical evolution, and
determined the value of ${v_\odot}$ at (11.1,12.24,7.25) km/s.}
For $R_{0}= (8.0 \pm 0.5)$ kpc~\cite{reid1993distance}, one obtains $v_{0} = 241 \pm 17$ km/sec from the motion of Sagittarius $A^{\star}$~\cite{green2003effect}.

Ref.~\cite{gillessen2009orbit} studied motions of stars orbiting Sagittarius $A^{\star}$ and measured the distance from the Sun to the Galactic center to be $R_{0} = 8.28\pm{0.33}$ kpc. Ref.~\cite{reid2014trigonometric} employed very long baseline interferometry techniques to obtain parallax and proper motion data of masers and found a consistent value at $R_{0} = 8.34\pm{0.16}$ kpc.
 More recently, ref.~\cite{eilers2019circular} derived the Galactic rotation curve by relying on the available 6-d phase-space information on a sample of red giants at Galactocentric radii in the range 5 $\lesssim$ R $\lesssim$ 25 kpc, combining data from the APOGEE DR14~\cite{majewski2017apache}, {\it Gaia} DR2~\cite{brown2018gaia}, 2MASS~\cite{skrutskie2006two}, and WISE~\cite{wright2010wide} surveys, and employing the Jeans equation to relate the
circular velocity to the number density, as well as the
radial and transverse velocity dispersions
of the selected stellar tracers.
Following the analysis in \cite{hogg2019spectrophotometric}, they precisely determined the distances to the stellar tracers under study, bringing the systematic uncertainties for their estimate of $v_{0}$ down to the $(2-5)\%$ level.
Resorting to their modeling function for the circular velocity curve,
 $v_{0}(R)= (229.0 \pm 0.2){\rm km/s} - (1.7 \pm 0.1){\rm km/s/kpc}(R-R_{0})$, we find that for a distance of $8.34\pm{0.16}$ kpc the resulting circular velocity equals $(228.6 \pm 0.34)$ km/s.
This is our recommendation to the experimental collaborations for substituting the canonical IAU-reported value of 220 km/s~\cite{kerr1986review}.

We end this sub-section with a brief discussion about the correlation between the astrophysical parameters $R_{0}$ and $v_{0}$~\cite{reid2009trigonometric, mcmillan2010uncertainty, reid2014trigonometric}. Such a correlation further implies a correlation between $v_0$ and $\rhoDM$~\cite{mccabe2010astrophysical}, as the latter quantity is a function of $R_0$.
Ref.~\cite{reid2014trigonometric} suggests that the two parameters are no longer strongly correlated and constrains the angular solar motion $\Omega_{0}$ with an accuracy of $\pm 1.4 \%$ at 30.57 $\pm$ 0.43 km/s/kpc for $R_{0} = 8.34\pm{0.16}$ kpc.
However, assuming that such a correlation between $R_{0}$ and $v_{0}$ persists, the spatial dependence of the DM density $\rhoDM$ is contingent upon the particular choice of the DM density profile. For an NFW profile, this implies that if we fix $\Omega_{0}$ and $R_{0}$ at the aforementioned values, allowing $v_{0}$ to vary by $\pm 30$ km/s/kpc would alter the DM density by $^{-17}_{+22} \%$.
This change in the DM density then proportionally affects the differential scattering rate, as can be seen in eq.~\ref{eq:diff_rate}. By referring to table~\ref{tab:SHMrate}, the reader can get a sense of how changes in $v_0$ impact $\rhoDM$, and by extension the scattering rate.

%%%
\subsection{Escape Velocity}
%%%
The Galactic escape velocity sets the threshold beyond which a test particle is no longer bound to the Galactic potential at the solar location, $\Phi (R_{0})$, and is defined as $\vesc = \sqrt{-2 \Phi (R_{0})}$.
One approach to estimating $\vesc$ is built upon sampling high-velocity halo stars and postulating a power law ansatz for the tail of their velocity distribution, truncated at the escape speed~\cite{leonard1990local}; this approach is insensitive to the shape of the rotation curve.
The distribution function characterizing these stars is assumed to be isotropic and in steady-state.
 Another approach that does not rely on the above hypotheses, uses kinematic, photometric, and theoretical constraints to build a dynamical mass model of the Milky Way, and estimates $\vesc$ through the resulting Galactic potential~\cite{mcmillan2011mass}.

The RAdial Velocity Experiment (RAVE) team employs the former methodology by studying spectroscopic stellar data, and performing simulations which assess the validity of the ans{\"a}tze used to model the distribution of the stellar velocities.
Their measurement of $\vesc=544 ^{+64}_{-46}$ km/s at 90$\%$ confidence level (C.L.)~\cite{smith2007rave} is currently the standard value for DM DD searches;
a more recent estimate by RAVE finds $\vesc=533 ^{+54}_{-41}$ at 90$\%$ CL~\cite{piffl2014rave}.
Note that the dark halo model implemented by the RAVE collaboration leads to correlations among $\vesc$ and the other local DM halo parameters~\cite{piffl2014rave}; therefore, one must take care to use consistent halo parameters when adopting the reported value of $\vesc$~\cite{lavalle2015making}, and not treat $\vesc$ as an independent quantity.

The value of $\vesc$ can also be extracted from the velocities of halo stars, and this has become a reliable method thanks to the improvements in the astrometry frontier. These extractions use a simple model for the tail of the velocity distribution of halo stars at a fixed radius, based on a truncated power law~\cite{leonard1990local}.
A recent analysis noted that the power-law scaling of the high-velocity tail of the stellar sample is contingent upon features of the stellar halo~\cite{deason2019local}, and concluded that there is a smaller prior on the scaling compared to previous analyses~\cite{smith2007rave,piffl2014rave,monari2018escape}. This new prior is appropriate for strongly eccentric orbits in the solar vicinity, and when applied to data from {\it Gaia} DR2, results in $\vesc=528 ^{+24}_{-25}$ km/s~\cite{deason2019local}. We use this value of $\vesc$ for our updated analysis.
\subsection{Earth Velocity}
\label{sub:earth}
%%%
The final parameter we will discuss in this section is the Galactic velocity of the Earth, $v_E$. This parameter enters DM DD calculations as one must calculate the rate in the reference frame of the Earth. Therefore, the VDF that appears in eq.~\ref{eq:eta_int} should be evaluated at the detector's frame; this is achieved by performing a Galilean boost to the galactic-frame velocity distribution, ${ f_{G}}(\vec{v})$, in the following manner
\begin{align}
f_{E}(\vec{v}) =  f_{G}(\vec{v} + {\vec{v}}_{E}(t)),
\end{align}
where the subscripts $G,E$ indicate the Galactic- and Earth-frame, respectively.

The Earth's velocity, $\vec v_E(t)=\vec{v}_{LSR} + \vec{v}_\odot + \vec{v}_{E}^{orb}(t)$ describes the motion of the Earth with respect to the galactic frame and is comprised out of three components: $ \vec{v}_{LSR} = (0, v_{0}, 0)$ that denotes the motion of the LSR with $v_{0}$ the circular velocity at the Solar radius (see $\S$\ref{sub:circ_v}), $\vec{v}_\odot = (U_{\odot}, V_{\odot}, W_{\odot})$ that corresponds to the Sun's peculiar motion velocity vector with respect to the LSR, and $\vec{v}_{E}^{orb}(t)$ for the Earth's orbital motion
around the Sun, defined as
$\vec{v}_{E}^{orb}(t) = v_{E}^{orb}\left [\hat{{e_{1}}} \rm{sin}\lambda(t) - \hat{{e_{2}}} \rm{cos}\lambda(t) \right ]$~\cite{gelmini2001weakly} with $v_{E}^{orb}=29.8$ km/s, while $\lambda$(t) is the solar ecliptic longitude, and $\hat{{e_{1}}} = (-0.0670, 0.4927, 0.8676)$, $\hat{{e_{2}}}= (-0.9931, -0.1170, 0.01032)$ are the two orthogonal unit vectors defining the Earth's plane at the time of the spring equinox and summer solstice, respectively, expressed here in galactic coordinates. In this work, we adopt a value of $\bar v_E=232$ km/s with a modulation of $\pm 15$ km/s over the course of the year~\cite{Lee:2013xxa, mccabe2014earth}.
This time-dependent revolution of the Earth around the Sun leads to an annual modulation of the scattering event rate, an effect that represents one of the most striking signatures in DM detection and provides a powerful handle in differentiating between signal and background~\cite{drukier1986detecting}.

\subsection{Summary of Astrophysical Parameters}
In figs.~\ref{fig:eta_panel} and~\ref{fig:v2f_panel}, we show the effects of varying the astrophysical parameters on the $\eta$ and $v^2 f(v)$ functions, respectively, for the different halo models discussed in this section.
We see that the choice of astrophysical parameters has a significant effect on the shape of $\eta$, with a mild variation among the halo models. As we will see in the next section, this effect has substantial consequences on the predicted rates for DM-electron scattering. Therefore, we advocate in favor of updating the benchmark values for the SHM parameters that appear throughout the bibliography of DM searches, and recommend a new set of values that can be adopted by the experimental groups that study the DM-electron scattering interactions.
Table \ref{tab:SHM_parameters} displays all the relevant galactic parameters used in our analysis, accompanied by the values broadly used in the literature to date ({\it current}), as well as more recent determinations of the parameters in question ({\it suggested}). In what follows, we quantify the effects that the variations of the astrophysical parameters have on the calculations of DM-electron scattering.
\FloatBarrier
\begin{table}[htbp]
	\begin{center}
% 		\addtolength{\tabcolsep}{4pt}
		  \onehalfspacing
		\begin{tabular}{ccc
			}
			\hline
			& current & suggested \\\hline
			{$v_0~[\text{km/s}]$} & {$220^{+60}_{-20}$ ~\cite{kerr1986review}}& {$228.6 \pm 0.34$ ~\cite{eilers2019circular}}\\
			{$\vesc~[\text{km/s}]$}&{$544^{+56}_{-94}$~\cite{smith2007rave}}& {$528^{+24}_{-25}$ ~\cite{deason2019local}}\\
			{$\rhoDM~[\text{GeV/cm}^3]$}&{$0.4$~\cite{Tanabashi:2018oca}} &{$0.46 ^{+0.07}_{-0.09}$ ~\cite{sivertsson2018localdark}} \\
			{$v_{E}~[\text{km/s}]$}& {$232 \pm 15$~\cite{Lee:2013xxa}}&{$232 \pm 15$~\cite{Lee:2013xxa}}\\
			{$R_{0}~[\text{kpc}]$}& {$8.0 \pm 0.5$~\cite{reid1993distance}}&{$8.34 \pm 0.16$ ~\cite{reid2014trigonometric}} \\
			\hline
		\end{tabular}
%		  \doublespacing
	\end{center}
	\caption{Measurements for the local circular speed $v_0$, escape speed $\vesc$, local DM density $\rhoDM$, Galactic Earth velocity $v_E$, and solar radius $R_0$. The central values in the middle column correspond to the canonical values for the parameters, in use by experimental groups so far; the central values in the last column are our recommendations to the experimental collaborations for future use.}
	\label{tab:SHM_parameters}
\end{table}
\FloatBarrier

%\newpage
%%%%%%%%%%%%%%%%%%%%%%%%%%%%%%%%%
\section{Results}
\label{sec:results}
%%%%%%%%%%%%%%%%%%%%%%%%%%%%%%%%%
In this section, we present the results for the rate and cross-section sensitivities for DM-electron scattering, from varying the SHM over its associated parameters. We present the detailed results for the Tsallis and empirical models in appendices~\ref{app:tsallis} and~\ref{app:empirical}, respectively, while showing comparisons among all the models considered in this analysis throughout the main text.
\FloatBarrier
\begin{table}[!htbp]
    \centering
\includegraphics[width=0.8\textwidth]{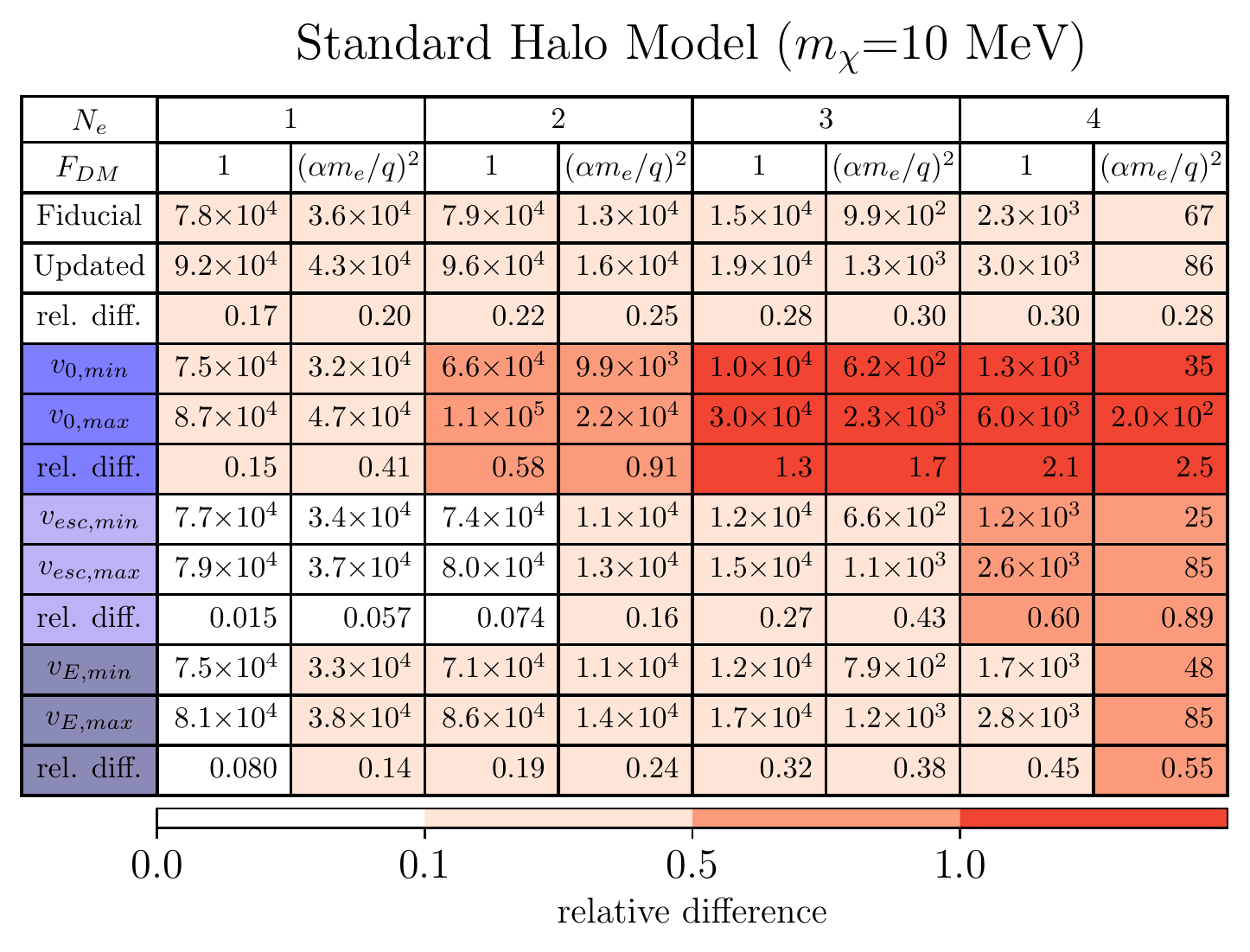}
\includegraphics[width=0.8\textwidth]{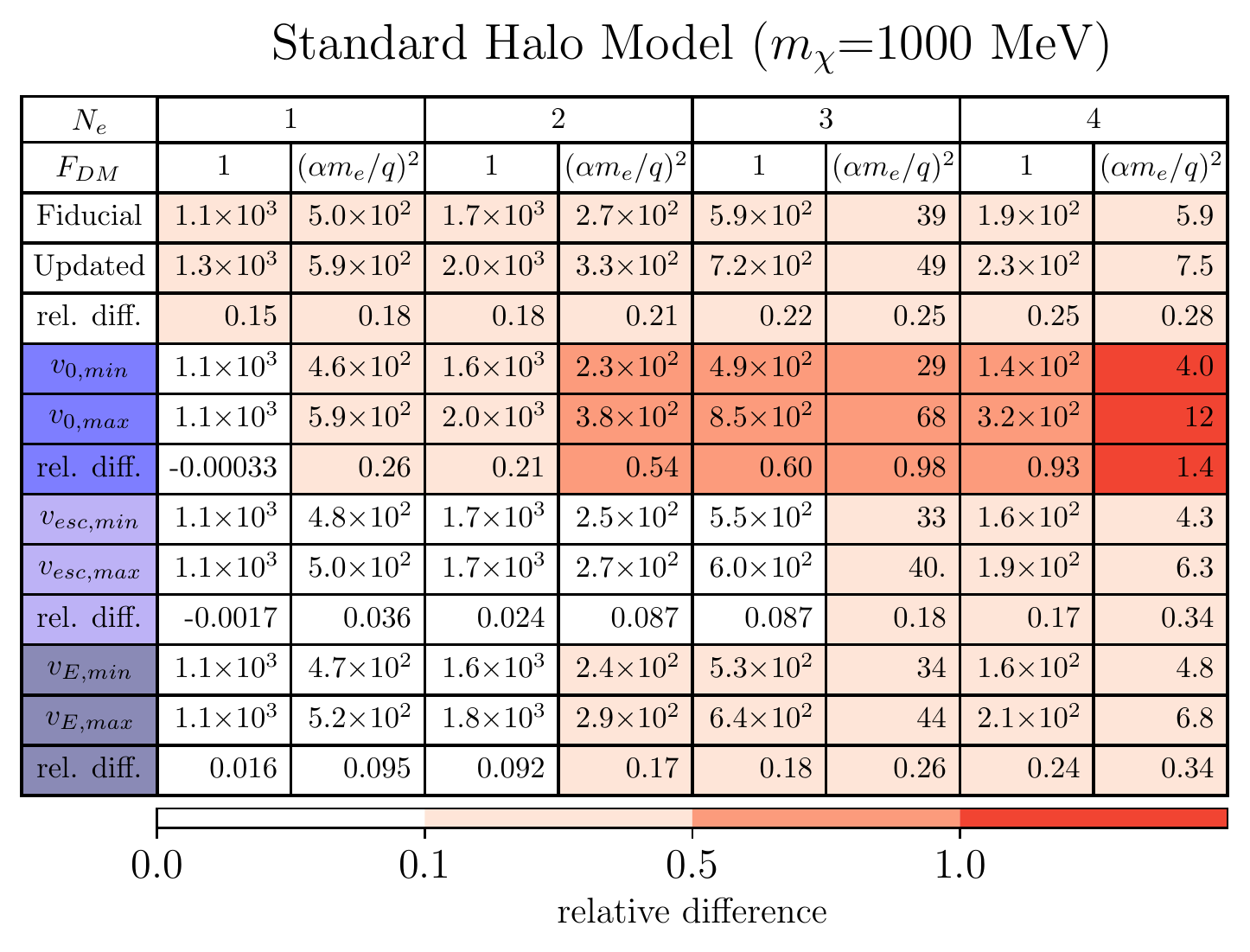}
    \caption{Expected number of events in 1 kg-year for various values of the SHM parameters $v_0, \vesc$, and $v_E$, as well as the relative difference between the minimum and maximum values for each parameter, for a given $N_e$ bin and DM form factor $F_{\rm DM}$. The top panel displays the results for $m_\chi=10$ MeV, and the bottom panel for $m_\chi=1$ GeV. The color scale indicates the size of the relative difference, with white (red) being the smallest (largest) relative difference.}
    \label{tab:SHMrate}
\end{table}
\FloatBarrier
\subsection{Standard Halo Model Rate and Cross Section}

We summarize the effects of varying over the parameters of the SHM on the predicted number of events in table~\ref{tab:SHMrate}, fixing the exposure at 1 kg-year and assuming a cross section of $\overline\sigma_e=10^{-37}$ cm$^2$. We calculate the number of events for each $N_e$ bin up to 4 (for the DM form factors $F_{\rm DM}=1$ and $F_{\rm DM}=(\alpha m_e / q)^2$) for each parameter combination, where e.g., $v_{0,\rm min}$ refers to $v_0$ at its minimum value with the other parameters held constant at their fiducial values. In each bin and for each parameter, the relative difference is defined by taking the difference between the rates for the maximum and minimum values of the underlying parameter, and dividing it by the fiducial rate, i.e.,

\begin{align}
	\mbox{rel. diff.} &= \frac{\mbox{Rate}_{\rm max} - \mbox{Rate}_{\rm min}}{\mbox{Rate}_{\rm fid}}.
\end{align}

From this table we see that $v_0$ gives the largest variance in the rate for these 4 bins for $m_\chi=10~(1000)$ MeV, ranging from $\sim 15\%~(0.03\%)$ for $N_e=1$ and $F_{\rm DM}=1$ to $\sim250\%~(140\%)$ for $N_e=4$ and $F_{\rm DM}=(\alpha m_e / q)^2$. The larger differences at low masses are to be expected, as low mass DM is more sensitive to the high-speed tail of the velocity distribution.
Similar tables that summarize how the count rates and cross sections are affected due to the variation of the underlying halo model for the Tsallis and empirical distributions can be found in appendices~\ref{app:tsallis} and~\ref{app:empirical}, respectively.

\begin{figure}[h]
    \centering
    \includegraphics[width=1\textwidth]{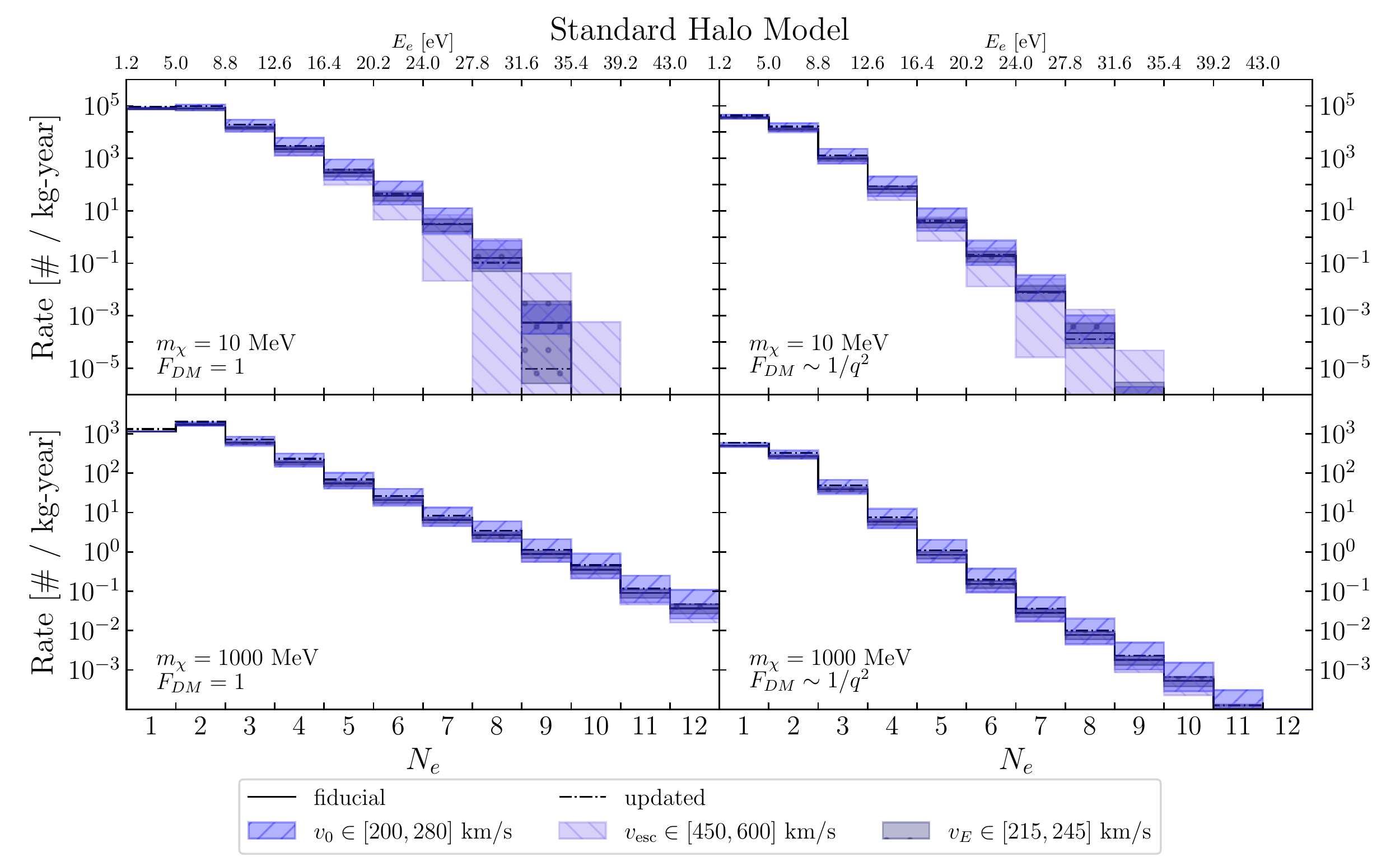}
    \caption{Rate spectra for the SHM at 1 kg-year exposure and $\sige=10^{-37}$ cm$^2$ for a range of halo parameters, as a function of the number of observed electrons, $N_e$, and electron energy $E_e$. The rate for the SHM with fiducial halo parameters is given by the black solid line, while the rate for the updated halo parameters is given by the black dash-dotted line. The top (bottom) panels show the rate for $m_\chi=10$ MeV ($m_\chi=1$ GeV), and the left (right) panels for $F_{\rm DM}=1$ ($F_{\rm DM}=(\alpha m_e/q)^2$).}
    \label{fig:SHMrate}
\end{figure}

In fig.~\ref{fig:SHMrate}, under the assumption of the SHM for the DM velocity profile, we present the spectrum of events as a function of the electron energy $E_e$, for two DM form factors ($F_{\rm DM} = 1$ and $F_{\rm DM}=(\alpha m_e / q)^2$) and two candidate masses ($m_{\chi} = 10$ MeV and 1 GeV). As expected, increasing the ionization threshold $N_e$ leads to a rate drop-off. However, for the lower-mass candidate at the high-energy regime, this fall-off is much more pronounced; lowering $\vesc$ significantly affects the rate, since for more energetic transitions to be realized, DM particles need to be moving much faster. Thus by decreasing $\vesc$, we effectively constrain the number of particles in the halo that are available to scatter. In all other cases we see that $v_0$ is leading the changes to the event rate, as is the case for the SHM's associated inverse mean speed in fig.~\ref{fig:eta_panel}.
Analogous plots that show how variations in the halo-model parameters impact the rates for the Tsallis and empirical distributions can be found in the appendices~\ref{app:tsallis} and~\ref{app:empirical}, respectively.

\begin{figure}[h]
    \centering
    \includegraphics[width=1\textwidth]{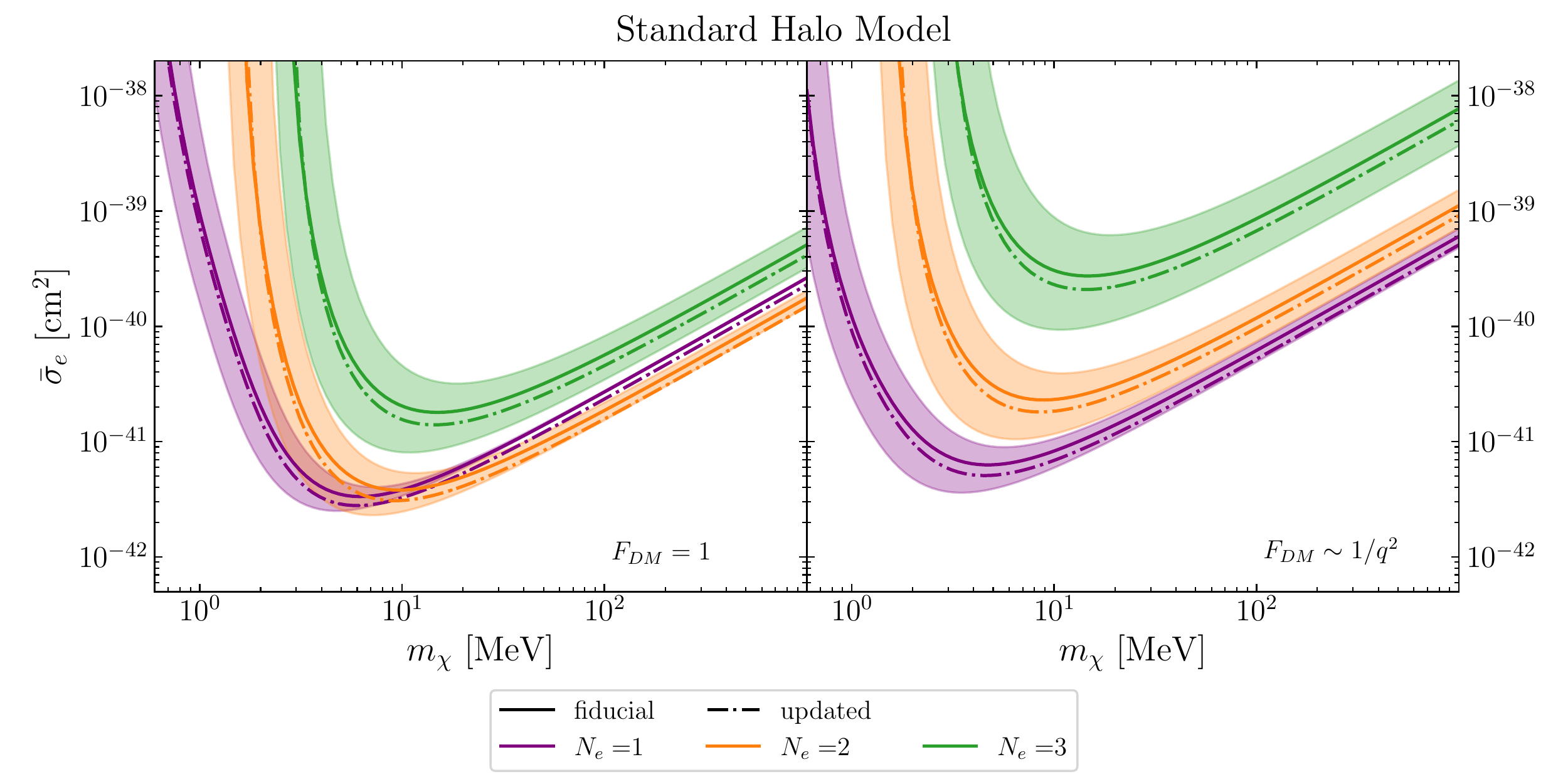}
    \caption{Projected 95$\%$ C.L. cross-section reach for the SHM, assuming 1 kg-year exposure and zero background events. The predictions for the $N_e=1,2,3$ bins are given in purple, orange, and green respectively. The solid curves are calculated using the fiducial values for the halo parameters, while the dash-dotted curves are calculated with the updated parameters. The bands denote the range of cross sections resulting from the variation of all the halo parameters within their allowed intervals.}
    \label{fig:sigmaSHM}
\end{figure}

In fig.~\ref{fig:sigmaSHM}, we show the projected sensitivity at 95\% C.L. ($\sim$3 events) to the DM-electron scattering cross section, $\sige$, as a function of the DM mass, $m_{\chi}$, setting the exposure at 1 kg-year and assuming zero background events. We present the resulting curves for the first three $N_e$ bins, which correspond to the purple, orange, and green lines, respectively. The color bands around each curve indicate the maximal and minimal differences in $\sige$, resulting from the variation of the SHM VDF across its parameters, $v_0$, $\vesc$, and $v_E$. Specifically, the lower edge of each band results from setting these three velocities at their maximum values, while the upper part corresponds to the minimum values of the velocities' respective ranges. When increasing the DM mass, this DD observable becomes more resilient to changes in the halo parameters, especially for the $N_e=1$ bin. In the lower end of the mass spectrum, changes in the halo-model parameters can change the lowest detectable mass by up to a factor of $\sim 2$. The updated parameters for the SHM generally result in an improved reach, primarily driven by the increase in $\rhoDM$. Although the SHM is most sensitive to variations in $v_0$, our recommendation for using more recent measurements shifts the central value of this parameter by only a few percent. In contrast, the updated $\rhoDM$ is $\sim 15\%$ larger; the effect of this larger value for the density is particularly noticeable at high masses, where the cross-section limits are the least sensitive to variations in the other halo-model parameters.
 Similar plots that show how the cross section is affected by the variation of the halo model across its associated parameters for the Tsallis and empirical distributions can be found in the appendices~\ref{app:tsallis} and~\ref{app:empirical}, respectively.

\subsection{Comparison Between Halo Models}
\label{sec:halocomp}
\begin{figure}[h]
    \centering
    \includegraphics[width=1\textwidth]{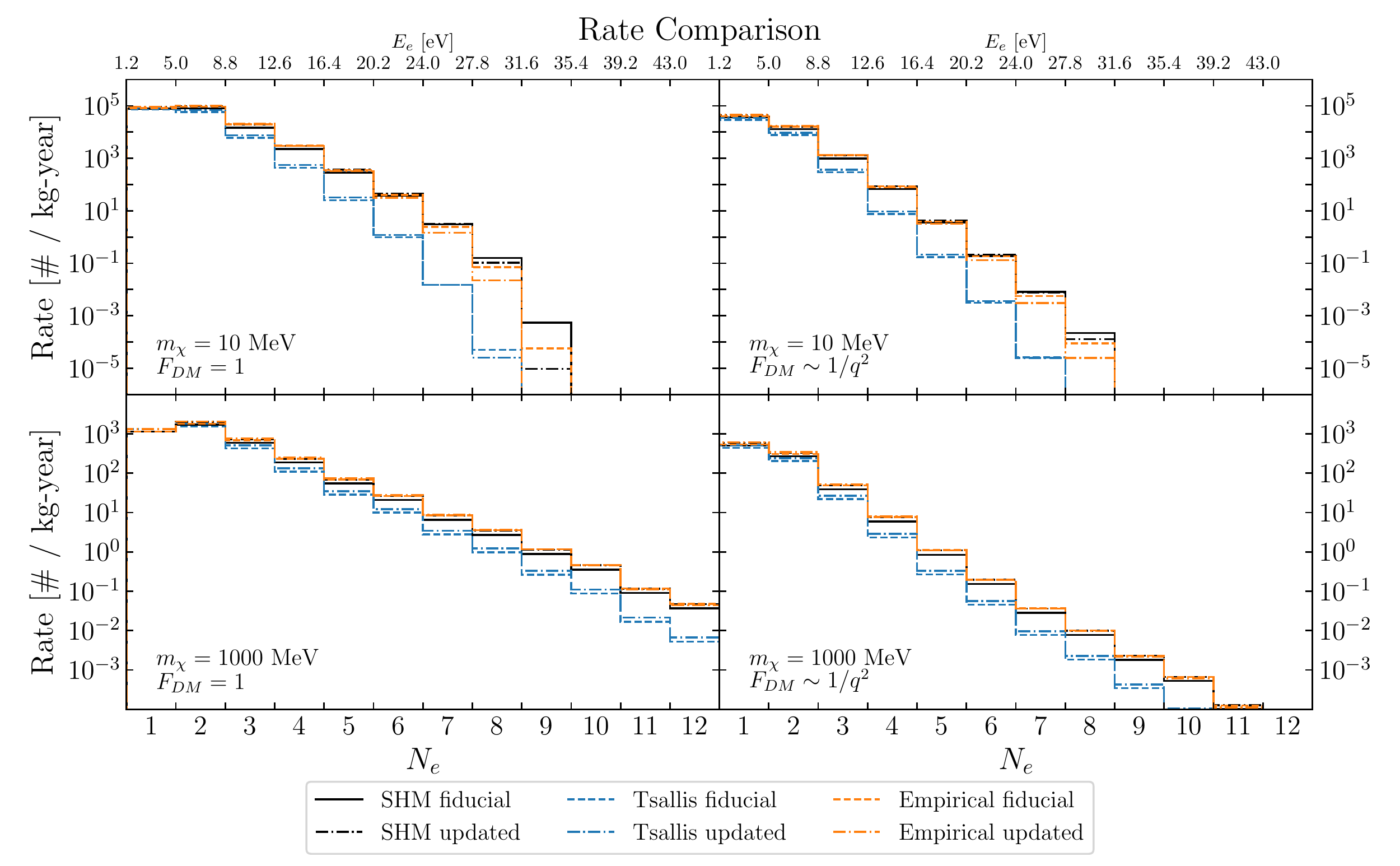}
    \caption{Comparison of the rate spectra for the SHM (black), Tsallis (blue), and empirical (orange) velocity distributions, assuming  $\overline\sigma_e=10^{-37}$cm$^2$ and 1 kg-year exposure. The solid/dashed (dash-dotted) curves correspond to the rates evaluated at their fiducial (updated) parameters. The top (bottom) panels show the rate for $m_\chi=10$ MeV ($m_\chi=1$ GeV), and the left (right) panels for $F_{\rm DM}=1$ ($F_{\rm DM}=(\alpha m_e/q)^2$). }
    \label{fig:ComboRate}
\end{figure}

In fig.~\ref{fig:ComboRate}, we compare the scattering event rates for the three VDFs considered in this work. We note that, as an overall trend, the empirical model tends to result in slightly higher rates than the SHM, while the Tsallis model yields drastically lower rates. This is especially true for the higher $N_e$ bins, where we are more sensitive to the high-speed tails of the VDFs. This trend can be traced back to the corresponding $\eta$ plots for the models in fig.~\ref{fig:eta_panel}, where for high $\vmin$ values, the Tsallis distribution results in a significantly lower $\eta$, and by extension to smaller rates.

The qualitative differences in the differential rates, for the three VDFs under study, call for a statistical analysis in pursuit of the minimum exposure at which one can differentiate among the underlying halo models. Note that such an analysis is only feasible if the DM mass and model are known; determining these requires a separate, independent measurement, the details of which are beyond the scope of this work. We proceed by assuming we know the DM mass and model. One can then perform a simple log-likelihood shape test to compare the event rates for the various velocity distributions. Following the procedure outlined in ref.~\cite{Porter:2008mc}, we find that the SHM and Tsallis distributions can be easily distinguished for exposures below a kg-year and DM masses in the range 1 MeV-1 GeV, whereas differentiating between the SHM fiducial and updated models requires much larger exposures, ranging from a few hundred gram-years for $m_\chi\simeq 10$ MeV to above 30 kg-years for $m_\chi\simeq 1$ GeV. These preliminary results motivate a more in-depth statistical analysis of DM-electron scattering data, which we leave for future work.

\begin{figure}[h]
    \centering
    \includegraphics[width=1\textwidth]{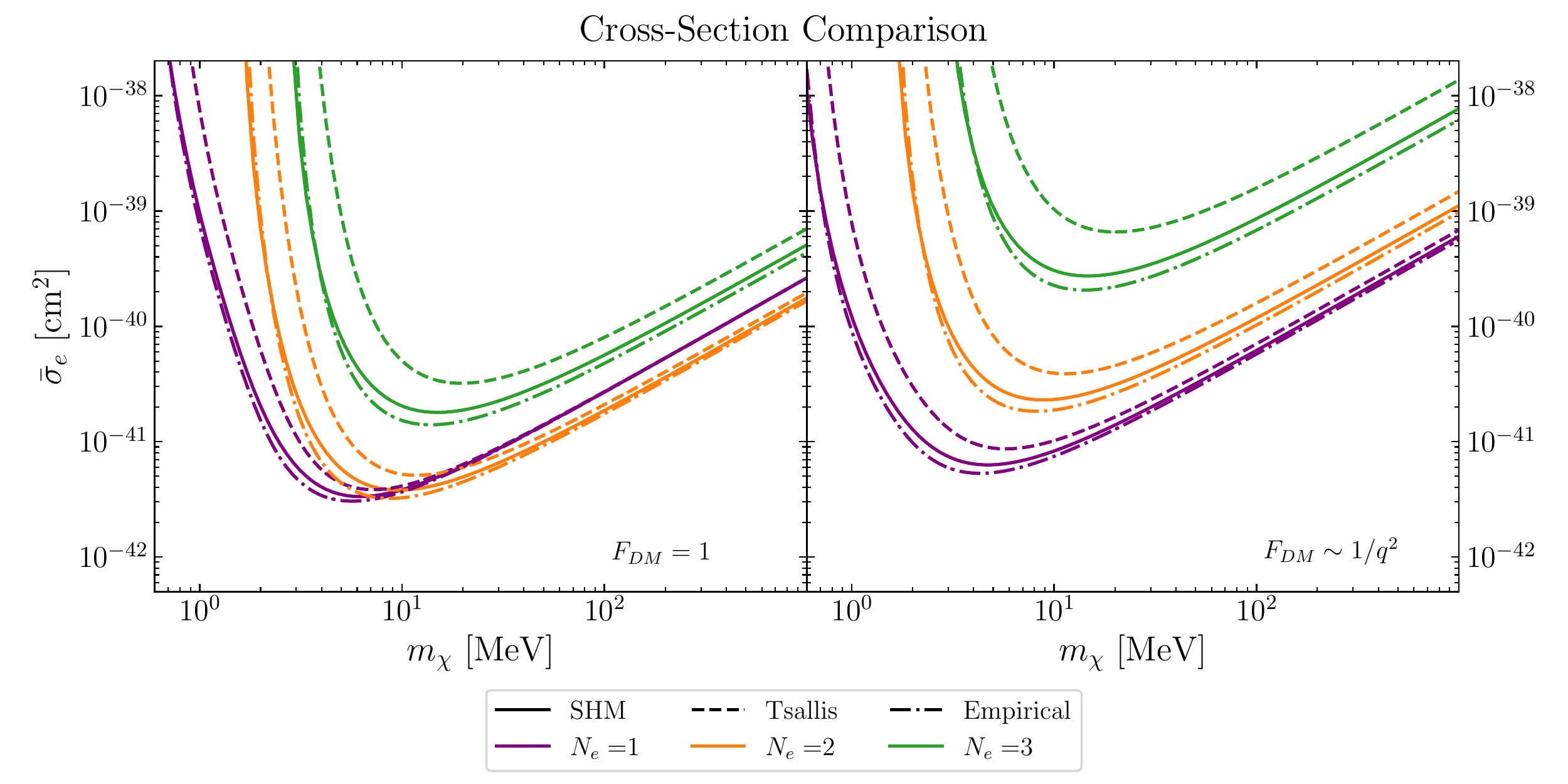}
    \caption{Cross-section comparison between the SHM (solid), Tsallis model (dashed), and empirical model (dash-dotted) for $N_e=1,2,3$ (purple, orange, green). All 95$\%$ C.L. curves are computed using the canonical values for the halo parameters ($v_0=220$ km/s, $\vesc=544$ km/s, $v_E=232$ km/s, and $p=1.5$).}
    \label{fig:sigmaCombo}
\end{figure}

In fig.~\ref{fig:sigmaCombo}, we compare the cross-section 95$\%$ C.L. projections for a kg-year exposure across the three VDFs under consideration. In this figure, the solid lines correspond to the SHM, dashed to the Tsallis model, and dash-dotted to the empirical model. We observe that in all cases, similarly to the rate, the Tsallis model results in a more limited reach, compared to the SHM and empirical model. For the most part, the empirical model exhibits a slightly better reach than the SHM, as confirmed from the first three $N_{e}$ bins of the rate plots as well. As in the SHM case, the cross section shows increased resilience to the choice of the VDF at higher masses, and especially for the $N_e=1$ bin.
Interestingly, for our choice of parameters, the empirical model closely follows the SHM.

\subsection{Annual Modulation of the Scattering Rate}

\begin{figure}[h]
    \centering
    \includegraphics[width=1\textwidth]{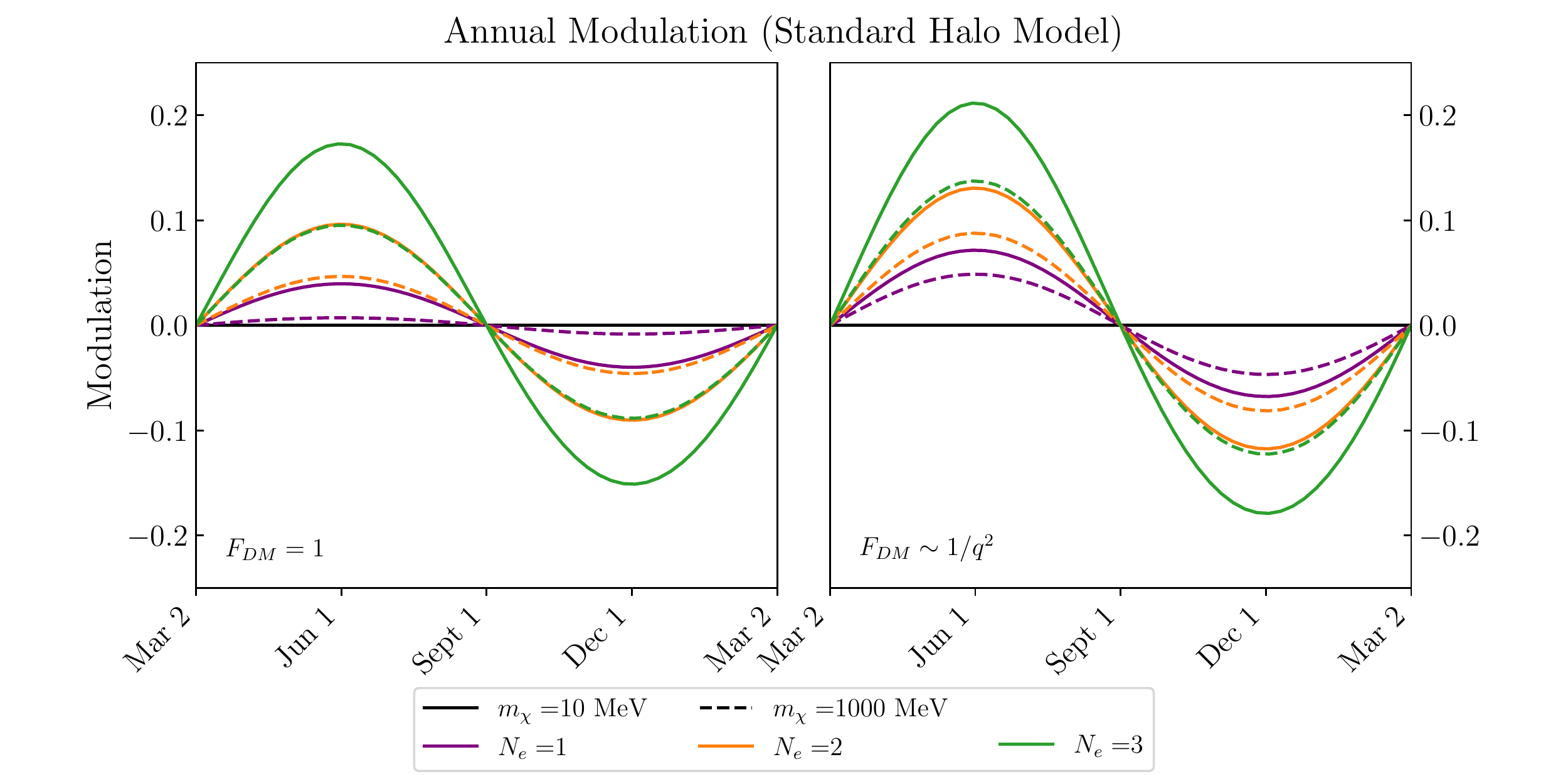}
    \includegraphics[width=1\textwidth]{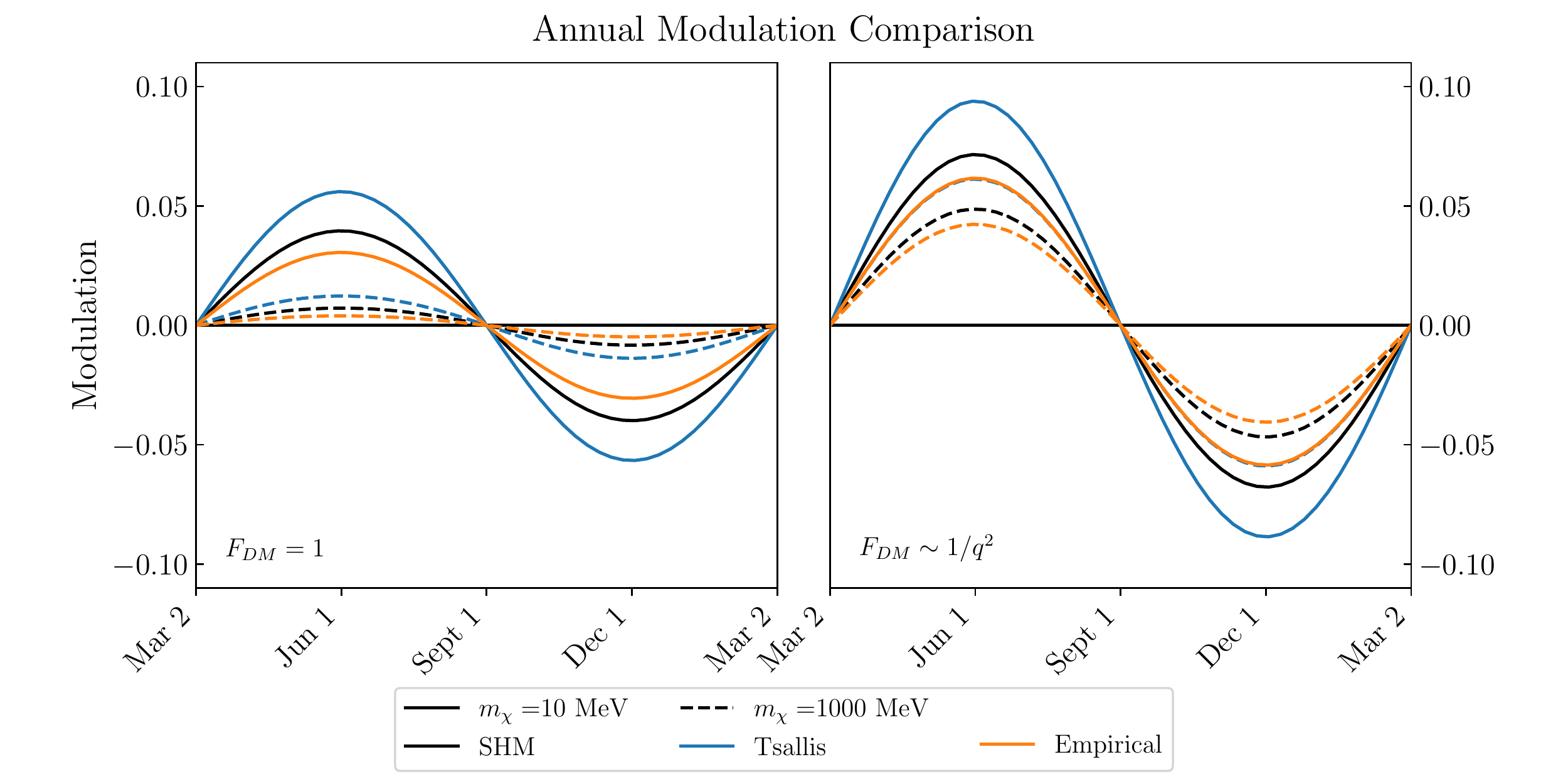}
    \caption{Annual modulation for $F_{\rm DM}=1$ (left) and $F_{\rm DM}\propto 1/q^2$ (right). The top panels shows the modulation for the SHM for $m_\chi=10$ MeV (solid), 1 GeV (dashed) and $N_e=1,2,3$ (purple, orange, green). The bottom panels show the modulation comparison among the SHM (black), Tsallis (blue), and empirical (orange) distributions for $N_e=1$.}
    \label{fig:ratetime}
\end{figure}

Due to the Earth's annual rotation around the Sun, its velocity relative to the Galactic center, $v_E$, evolves in time, as a result of which the count rate experiences an annual modulation. In this section, we briefly discuss the behavior of this annual modulation for the three considered halo models; a more detailed treatment of the annual modulation of the DM-electron scattering rate under the assumption of the SHM can be found in ref.~\cite{Essig:2015cda}.
The annual modulation of the scattering rate is defined as
\begin{align}
    \mbox{Modulation}(N_e, t) &= \frac{\mbox{Rate}(N_e, v_E(t)) - \mbox{Rate}(N_e, \bar{v}_E)}{\mbox{Rate}(N_e, \bar{v}_E)},
\end{align}
where $\bar{v}_E = 232$ km/s is the fiducial average $v_E$ value. The other halo parameters are set to their respective fiducial values.

In fig.~\ref{fig:ratetime} we show the annual modulation of the rate as a function of time. The plots in the top row correspond to the resulting modulation for the SHM for the $N_e=$ 1, 2, 3 bins. The bottom panels present the comparison among the SHM (black), Tsallis (blue) and empirical (orange) distributions for $N_e=$ 1. The modulation is calculated for two candidate masses, $m_\chi=10$ MeV (solid) and $m_\chi=1$ GeV (dashed), and two form factors $F_{\rm DM}=1$ (right) and $F_{\rm DM}=(\alpha m_e/q)^2$ (left).
 As the plots in fig.~\ref{fig:SHMrate} suggest, varying $v_E$ has a larger effect toward the higher $N_{e}$ bins, which are in general more susceptible to variations of the halo model, with the effect being more pronounced for the DM form factor proportional to $1/q^2$. These effects manifest themselves in the case of the annual modulation of the rate as well, as it is evident from the high-$N_{e}$ behavior of the modulation for the SHM, and the resulting modulation for all the models for the $F_{\rm DM}=(\alpha m_e/q)^2$ form factor. As emphasized earlier, larger DM masses tend to be more robust to changes in the astrophysical parameters, as verified by the lower annual modulation resulting for the $m_{\chi} = 1$ GeV candidate mass.

Comparing between the three VDFs, we see that the Tsallis model results in the greatest modulation, while the empirical model results in the smallest. This behavior is justified by the fact that the Tsallis velocity distribution is overall more sensitive to changes across its associated parameters, followed by the SHM and the empirical distribution. Since changes in $v_E$ are effectively changes in velocity for the calculation of the inverse mean speed $\eta$ (see eqn.~\ref{eq:eta_int}), this behavior propagates into the event rates resulting for each model.

%%%%%%%%%%%%%%%%%%%%%%%%%%%%%%%%%
\section{Discussion and Conclusions}
\label{sec:conclusions}
%%%%%%%%%%%%%%%%%%%%%%%%%%%%%%%%%

In this work, we quantify the dependence of the observables of DM-electron scattering, namely the differential scattering rate and the cross section, on the uncertainties of the underlying DM halo velocity distribution. We consider three halo-models: the SHM, the Tsallis model, and an empirical model. We study the effects of varying the astrophysical parameters within each model, and calculate the projected event rates and cross-section bounds. In addition, we give a brief overview of the measurements of the relevant astrophysical parameters and propose updated values for $v_0, \vesc,$ and $\rhoDM$.

Ultimately, we find that the empirical model sets the strongest constraints, while the Tsallis model sets the weakest. When varying the halo models across the relevant astrophysical parameters, we find that $v_0$ is the most influential parameter in the cases of the SHM and Tsallis distributions, while $p$ is the dominant parameter for the empirical model.
For $m_\chi=10~(1000)$ MeV, the variation in $v_0$ leads to changes in the rate ranging from $\sim 15\%~(0.03\%)$ for $N_e=1$ and $F_{\rm DM}=1$ to $\sim 300\%~(140\%)$ for $N_e=4$ and $F_{\rm DM}=(\alpha m_e / q)^2$. In the empirical model, the variation in $p$ leads to changes in the rate ranging from $\sim 5\%~(1\%)$ for $N_e=1$ and $F_{\rm DM}=1$ to $\sim 500\%~(240\%)$ for $N_e=4$ and $F_{\rm DM}=(\alpha m_e / q)^2$.
Overall, the empirical model varies the most within the allowed intervals for its parameters, while the SHM varies the least. Across all three models we see a pattern according to which, the direct detection observables depend more on our choice of VDF and its associated parameters for the lower DM masses, higher threshold energies ($N_e$), and $F_{DM} \sim 1/q^2$, while the choice of the VDF has a weaker effect when decreasing the DM mass, lowering the ionization threshold $N_e$ and setting $F_{DM} = 1$. For all three models, updating the canonical value of the local DM density with a higher value, leads to a stronger bound on the projected cross section.

The sensitivity of DM-nuclear scattering to astrophysical uncertainties is well-known, and must be taken into account when interpreting the results of the direct detection experiments (see {\it e.g.}~\cite{Benito:2020lgu}). Our results quantify the sensitivity of DM-{\it electron} scattering observables to the astrophysical parameters, and highlight the importance of extracting accurate measurements for these parameters, aiming at a consistent interpretation of the relevant experimental efforts. Several proposals account for these astrophysics-driven uncertainties in the case of DM-nuclear scattering, ranging from marginalizing over astrophysical uncertainties~\cite{Pato:2012fw}, to presenting the results in a halo-independent manner~\cite{fox2011interpreting, Fox:2010bz,gondolo2012halo,DelNobile:2013cta,Bozorgnia:2013hsa, fox2014taking}. We emphasize that similar techniques can be developed and applied in the context of DM-electron scattering interactions and are worthy of a dedicated study.

Once data from ongoing and upcoming DD experiments start flowing in, the emergence of a positive signal would mark the onset of the DM discovery mode.
Such a development would necessitate/expand the design of statistically-robust analyses, allowing one to accurately reconstruct the properties of the DM particle, through the study of the available DD data~\cite{Pato:2012fw}. Once such analyses would be in place, one would be in position to ask how well the DM VDF can be determined, by performing various statistical tests with the potential to pin down the underlying DM halo model. Preliminary analyses along these lines show great promise for DM-electron scattering~\cite{Buch:2020xyt}, and this direction warrants further exploration.

\section*{Acknowledgements}
We would like to thank Kyle Cranmer, Ian Lewis, and David Strom for many useful discussions about statistics. We would also like to thank Tim Cohen and Paddy Fox for comments on an earlier version of this manuscript. This work is supported in part by the NSF CAREER-1944826-PHY grant.

\begin{appendices}

\section{Tsallis Rate and Cross-Section Results}\label{app:tsallis}
\begin{figure}[htbp]
    \centering
    \includegraphics[width=1\textwidth]{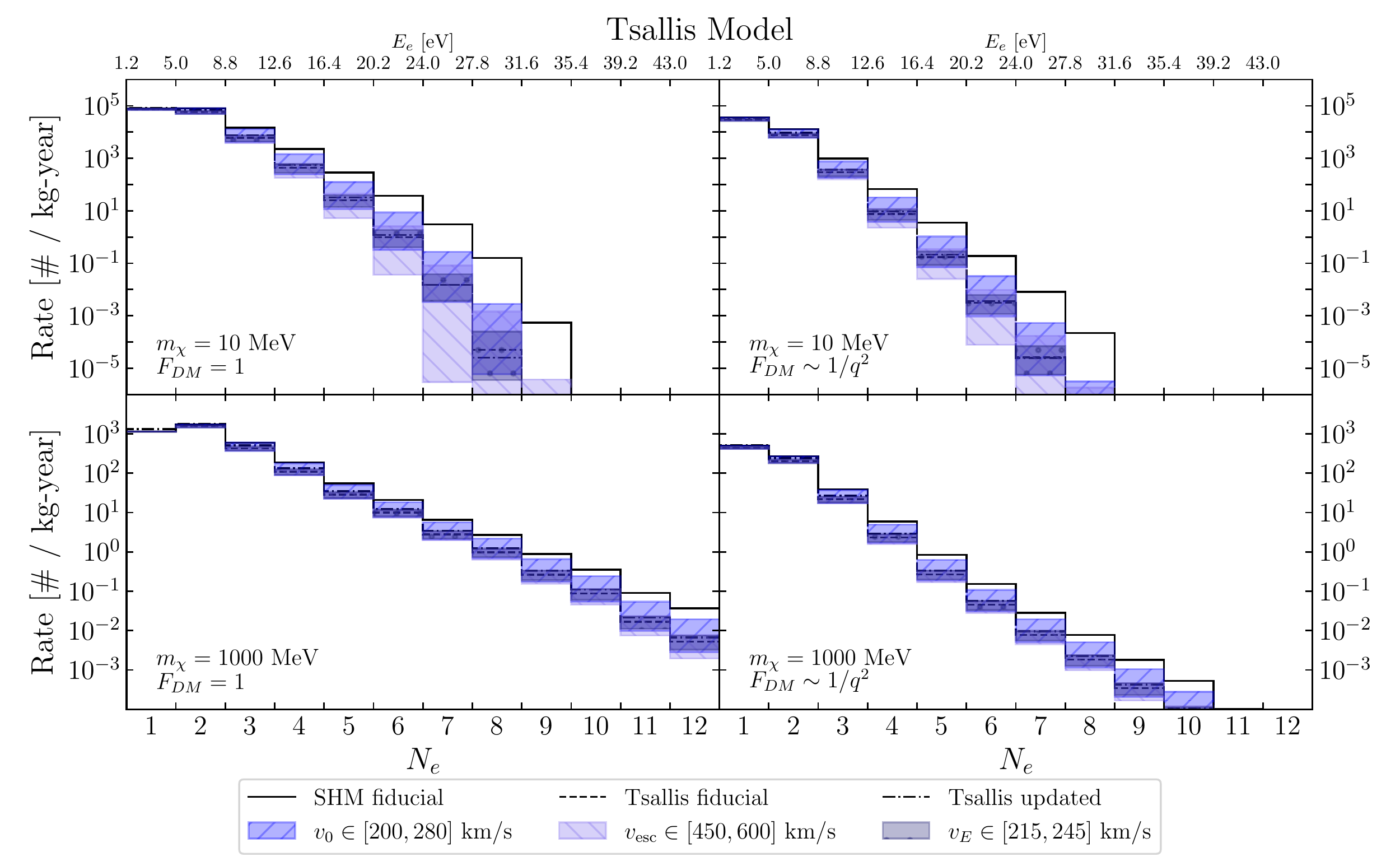}
    \caption{Rate spectra for the Tsallis model at 1 kg-year exposure and $\sige=10^{-37}$ cm$^2$ for a range of halo parameters, as a function of the number of observed electrons, $N_e$, and electron energy $E_e$. The rate for the Tsallis distribution with fiducial halo parameters is given by the black dashed line, while the rate for the updated halo parameters is given by the black dash-dotted line. For comparison, we show the rate for the SHM with fiducial halo parameters in the black solid line. The top (bottom) panels show the rate for $m_\chi=10$ MeV ($m_\chi=1$ GeV), and the left (right) panels for $F_{\rm DM}=1$ ($F_{\rm DM}=(\alpha m_e/q)^2$).}
    \label{fig:Tsarate}
\end{figure}
\begin{figure}[!h]
\centering
\includegraphics[width=1\textwidth]{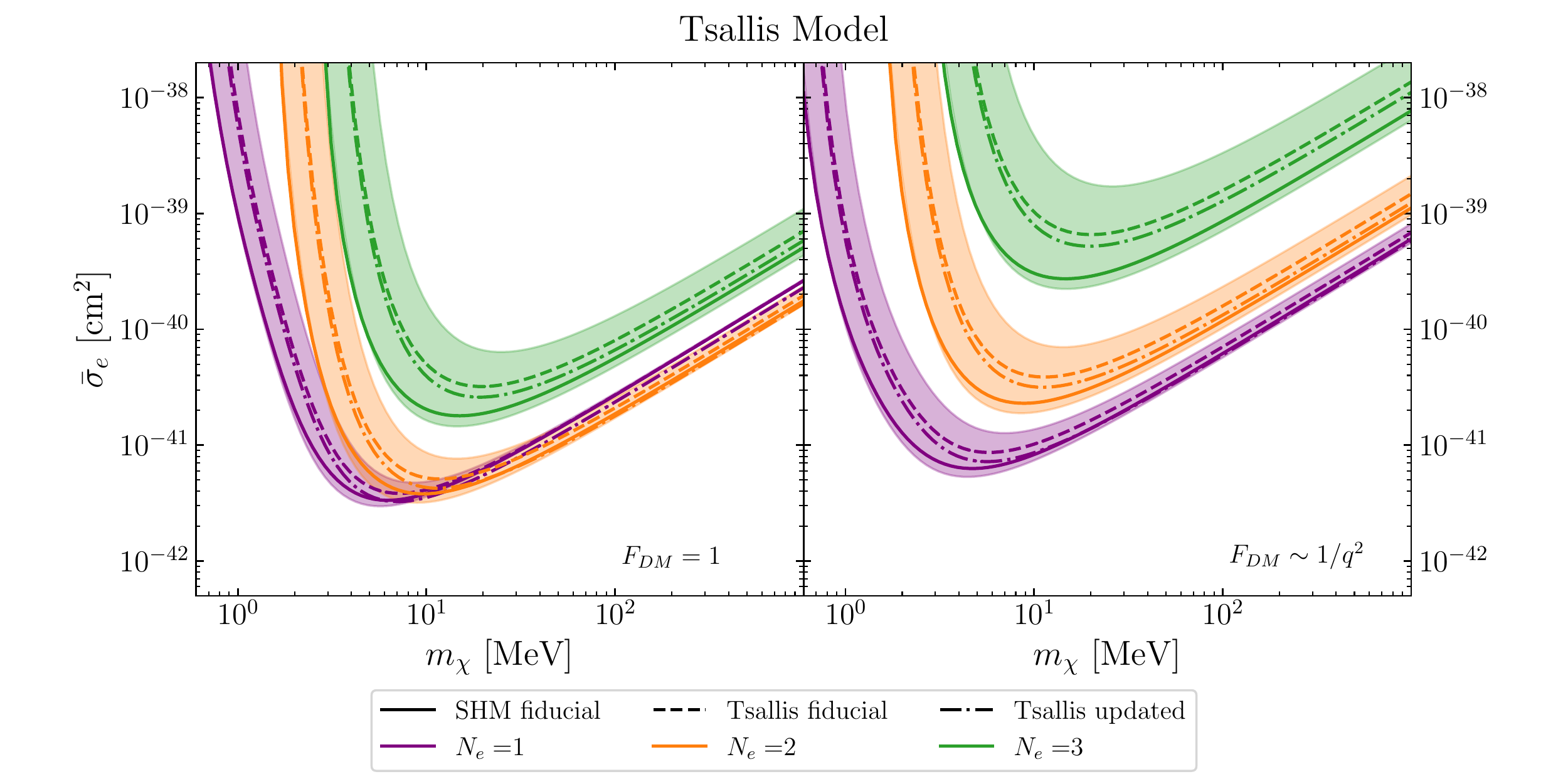}
\caption{Projected 95$\%$ C.L. cross-section reach for the Tsallis model, assuming 1 kg-year exposure and zero background events. The predictions for the $N_e=1,2,3$ bins are given in purple, orange, and green respectively. The dashed curves are calculated using the fiducial values for the halo parameters, while the dash-dotted curves are calculated with the updated parameters. The bands denote the range of cross sections resulting from the variation of all the halo parameters within their allowed intervals. The solid curves for the SHM, evaluated at its fiducial halo parameters, are shown for comparison.}
\label{fig:sigmaTsa}
\end{figure}

\begin{figure}[!h]
    \centering
    \includegraphics[width=1\textwidth]{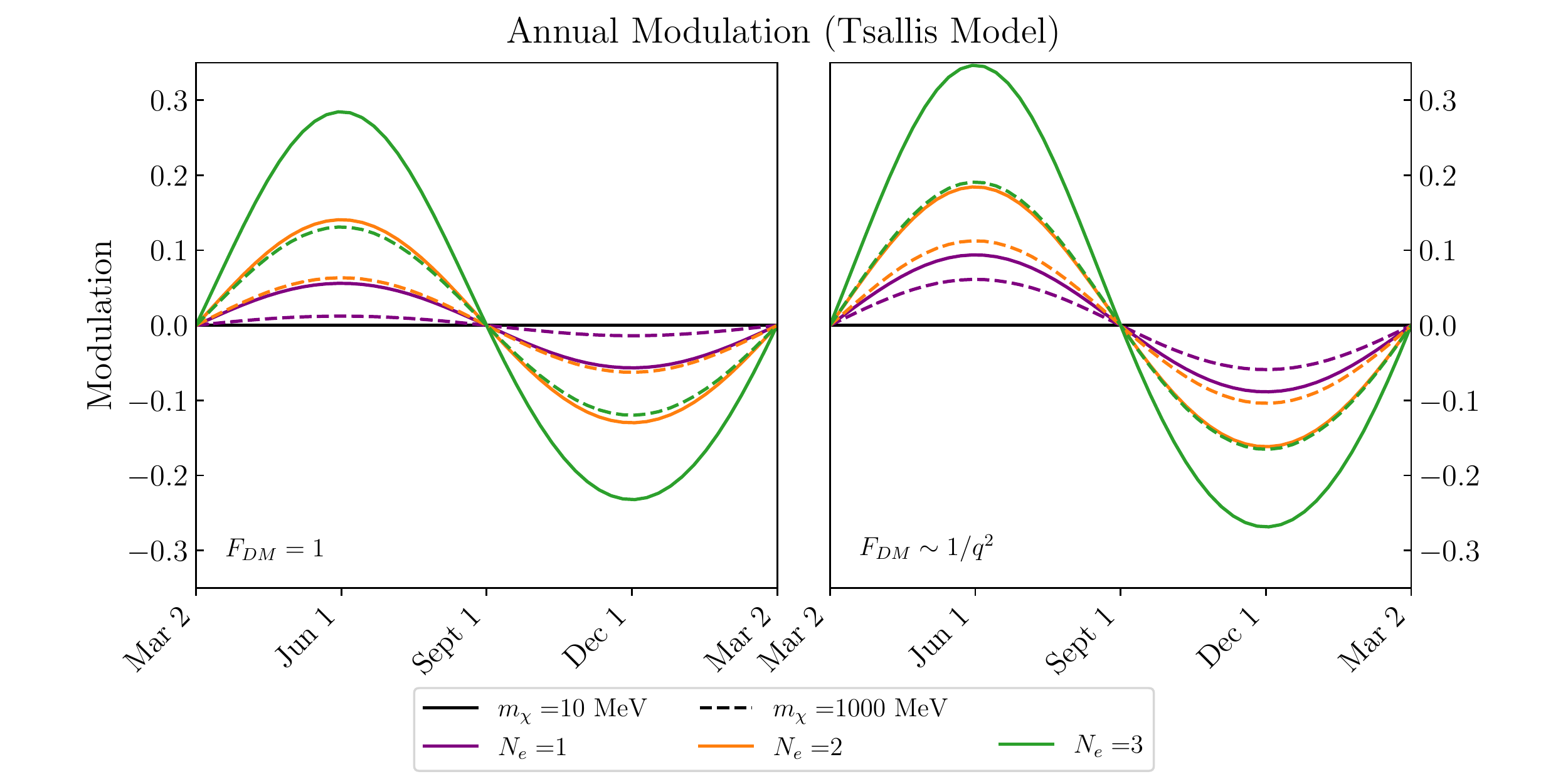}
    \caption{Annual modulation of the scattering rate for the Tsallis distribution for $F_{\rm DM}=1$ (left) and $F_{\rm DM}\propto 1/q^2$ (right) for two candidate masses $m_\chi=10$ MeV (solid), 1 GeV (dashed) and $N_e=1,2,3$ (purple, orange, green).}
    \label{fig:ratetime_Tsa}
\end{figure}
In this appendix, we present the results for the Tsallis model in greater detail. In fig.~\ref{fig:Tsarate}, we show the spectrum of events as a function of the electron energy $E_e$, for two DM form factors ($F_{\rm DM} = 1$ and $F_{\rm DM}=(\alpha m_e / q)^2$) and two candidate masses ($m_{\chi} = 10$ MeV and 1 GeV). The Tsallis distribution exhibits higher sensitivity to variations across its parameters, and especially $\vesc$, as we move toward higher $N_e$ bins and lower DM masses. The Tsallis model appears to be more sensitive when varied over its parameters, compared to the SHM.

In fig.~\ref{fig:sigmaTsa}, we see the predicted 95$\%$ C.L. cross-section reach, assuming 1 kg-year exposure and no background events. Here we notice that the Tsallis model once again shows larger variation when varied across its parameters, compared with the SHM case, and when evaluated at the maximal values allowed for its associated parameters it can reach the fiducial SHM limits. As in the case of the SHM, we see that increases in DM mass lead to increased resistance to changes in the halo-model parameters, while increases in $N_e$ make the cross section more susceptible to such changes.

In fig.~\ref{fig:ratetime_Tsa}, we show the annual modulation for the Tsallis model. This distribution results in a larger annual modulation comparatively with the SHM. Like for the latter model, larger DM masses result in smaller modulation, as does the choice of the form factor for the heavy mediator, $F_{DM} = 1$.

\section{Empirical Rate and Cross-Section Results}\label{app:empirical}
\begin{figure}[!h]
    \centering
    \includegraphics[width=1\textwidth]{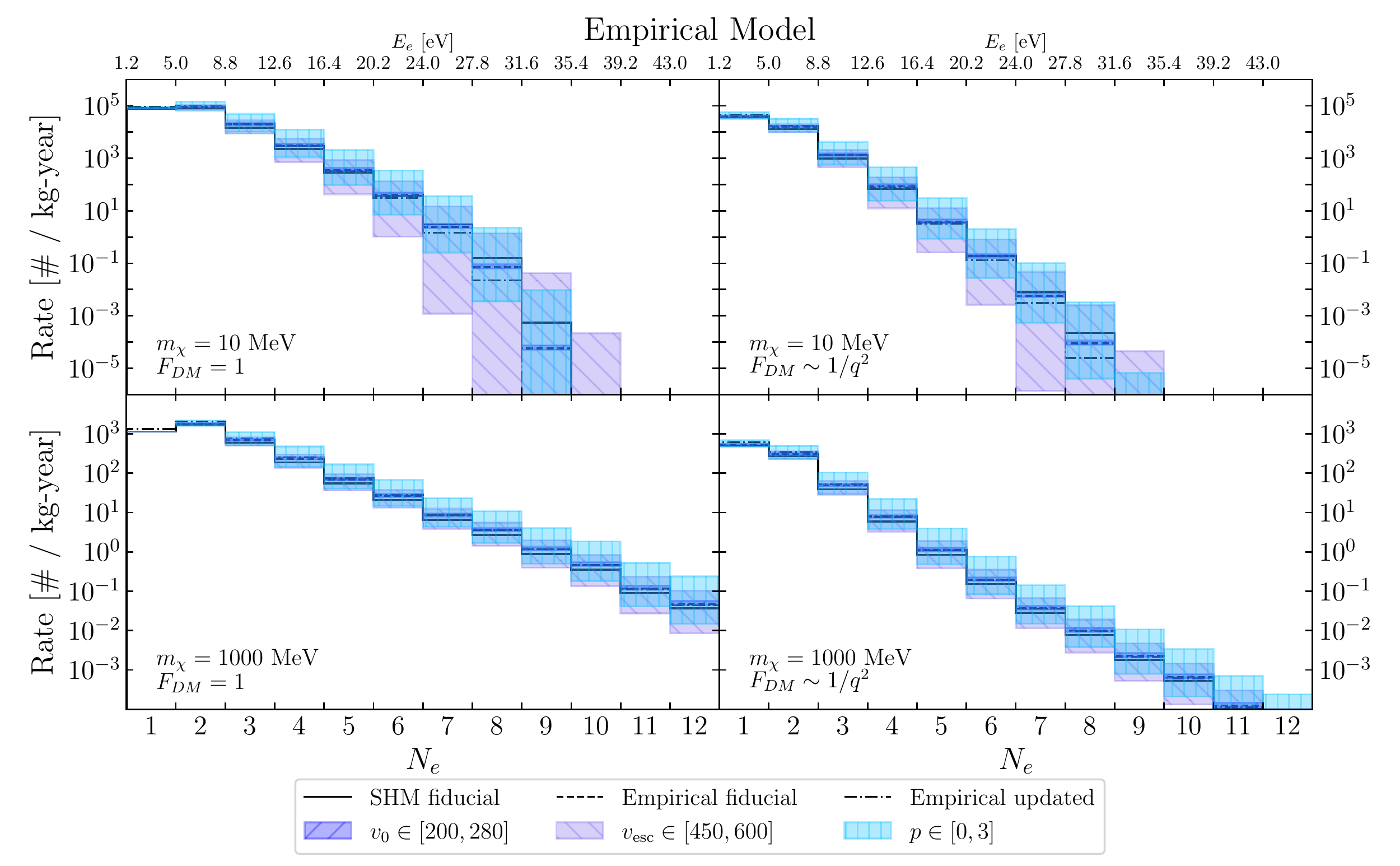}
    \caption{Rate spectra for the empirical model at 1 kg-year exposure and $\sige=10^{-37}$ cm$^2$ for a range of halo parameters, as a function of the number of observed electrons, $N_e$, and electron energy $E_e$. The rate for the empirical distribution with fiducial halo parameters is given by the black dashed line, while the rate for the updated halo parameters is given by the black dash-dotted line. For comparison, we show the rate for the SHM with fiducial halo parameters in the black solid line. The top (bottom) panels show the rate for $m_\chi=10$ MeV ($m_\chi=1$ GeV), and the left (right) panels for $F_{\rm DM}=1$ ($F_{\rm DM}=(\alpha m_e/q)^2$).}
    \label{fig:MSWrate}
\end{figure}

In this appendix, we present the results for the empirical model in more detail. In fig.~\ref{fig:MSWrate}, we show the spectrum of events as a function of the electron energy $E_e$, for two DM form factors ($F_{\rm DM} = 1$ and $F_{\rm DM}=(\alpha m_e / q)^2$) and two candidate masses ($m_{\chi} = 10$ MeV and 1 GeV). As we saw previously for the SHM and the Tsallis distributions, the empirical model shows more sensitivity when varied across its parameters as we increase the ionization threshold $N_e$ and decrease the DM mass, however now $p$ is the strongest source of variance in \textit{all} cases.

\begin{figure}[!h]
\centering
\includegraphics[width=1\textwidth]{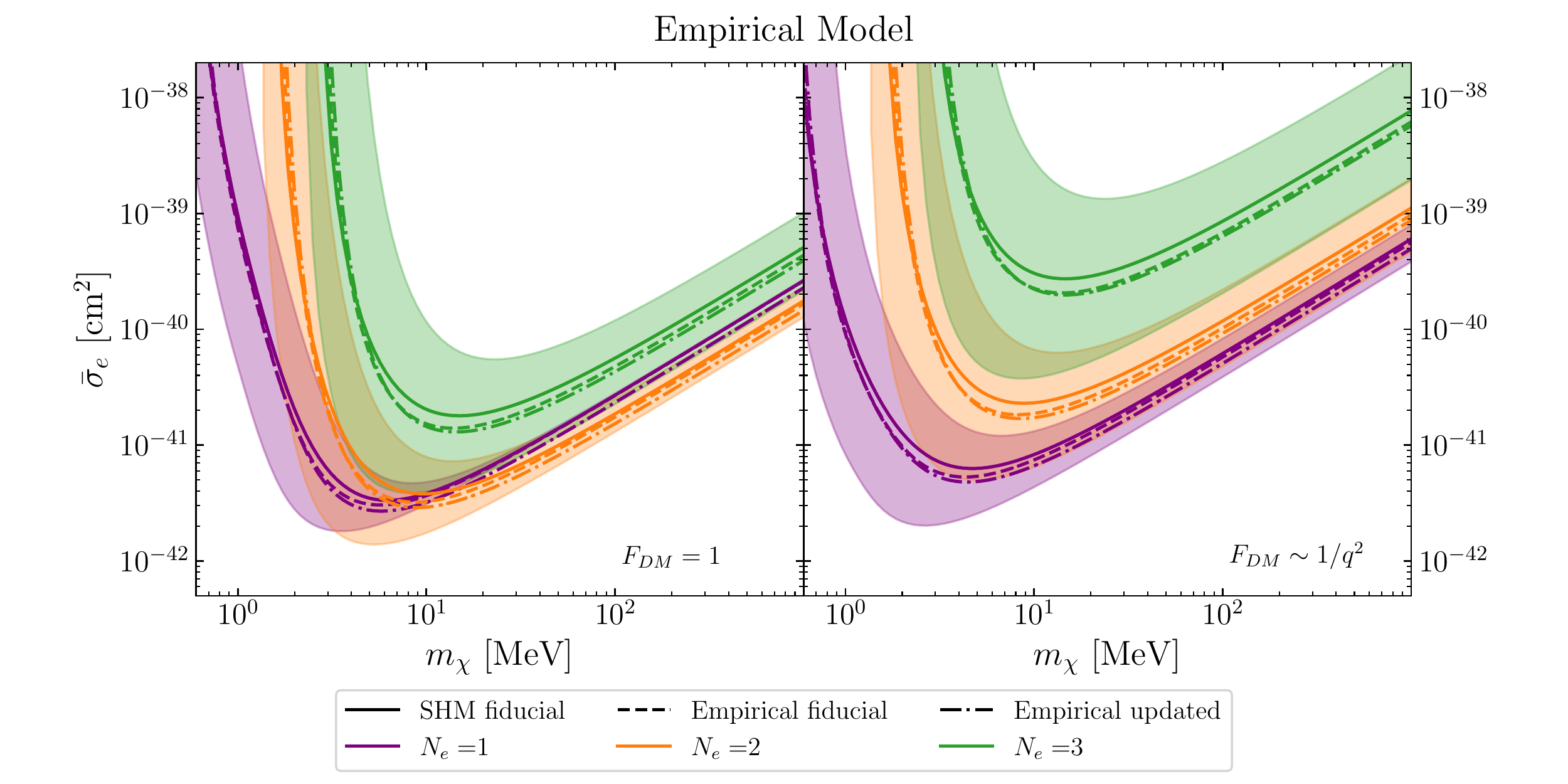}
\caption{Projected 95$\%$ C.L. cross-section reach for the empirical model, assuming 1 kg-year exposure and zero background events. The predictions for the $N_e=1,2,3$ bins are given in purple, orange, and green respectively. The dashed curves are calculated using the fiducial values for the halo parameters, while the dash-dotted curves are calculated with the updated parameters. The bands denote the range of cross sections resulting from the variation of all the halo parameters within their allowed intervals. The solid curves for the SHM, evaluated at its fiducial halo parameters, are shown for comparison.}
\label{fig:sigmaMSW}
\end{figure}

In fig.~\ref{fig:sigmaMSW}, we see the predicted 95$\%$ C.L. cross-section reach, fixing the exposure at 1 kg-year and assuming zero background events. Here, the top edges of the color bands result from substituting the minimum allowed values for $v_0$, $\vesc$, and $v_E$, while we have substituted the \textit{maximum} value for the $p$ exponent of the distribution, hence increasing $p$ would lead to \textit{weaker} bounds. Once again we see that the empirical model is affected more when varied across its parameters compared with the SHM and Tsallis distributions, and can well exceed the limits that could be set by either, depending on our choice of $p$. Results from the Eris simulation suite, that takes into account baryonic feedback effects, suggest a value for $p$ ($p$=3) greater than the corresponding value inferred by its DM-only counterpart ($p$=1.5); the choice of the larger value would result in weaker bounds for the empirical model.

\begin{figure}[!h]
    \centering
    \includegraphics[width=1.0\textwidth]{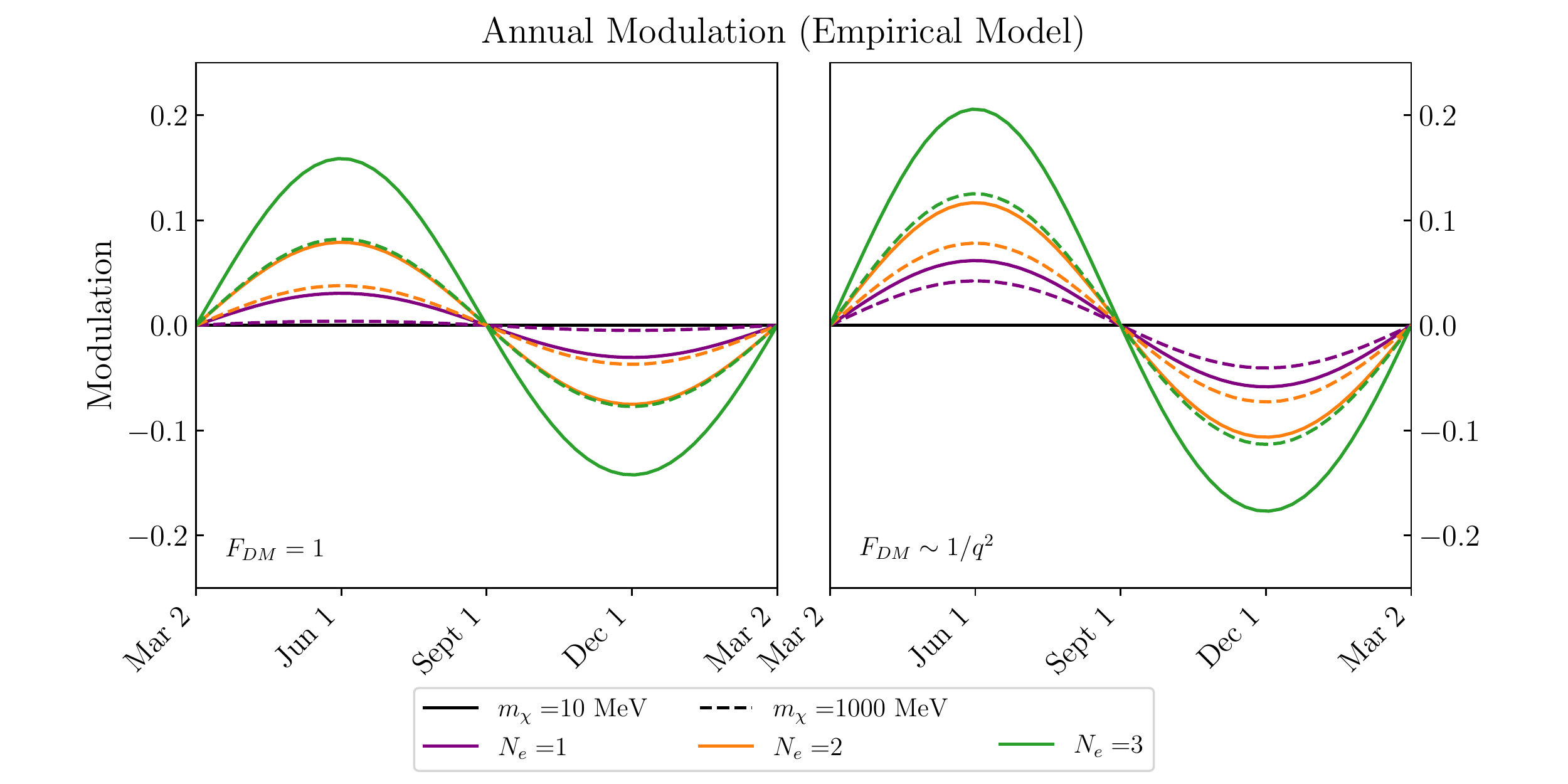}
    \caption{Annual modulation of the scattering rate for the empirical distribution for $F_{\rm DM}=1$ (left) and $F_{\rm DM}\propto 1/q^2$ (right) for two candidate masses $m_\chi=10$ MeV (solid), 1 GeV (dashed) and $N_e=1,2,3$ (purple, orange, green).}
    \label{fig:ratetime_MSW}
\end{figure}

In fig.~\ref{fig:ratetime_MSW}, we show the annual modulation of the count rate for the empirical model. As we saw in the two bottom plots in fig.~\ref{fig:ratetime}, the empirical model in general has a smaller modulation than the SHM, and follows the same trends as the SHM when it comes to the choice of the DM mass and form factor. Namely, larger masses result in smaller modulation, while a light mediator ($F_{DM} \sim 1/q^2$) would result in larger modulation. Although changes to the $p$ parameter affect the empirical model strongly, we see that variance in $v_E$ plays a minor role, as is the case for the other two velocity profiles.

\newpage

\FloatBarrier
\begin{table}[H]
    \centering
\includegraphics[width=0.8\textwidth]{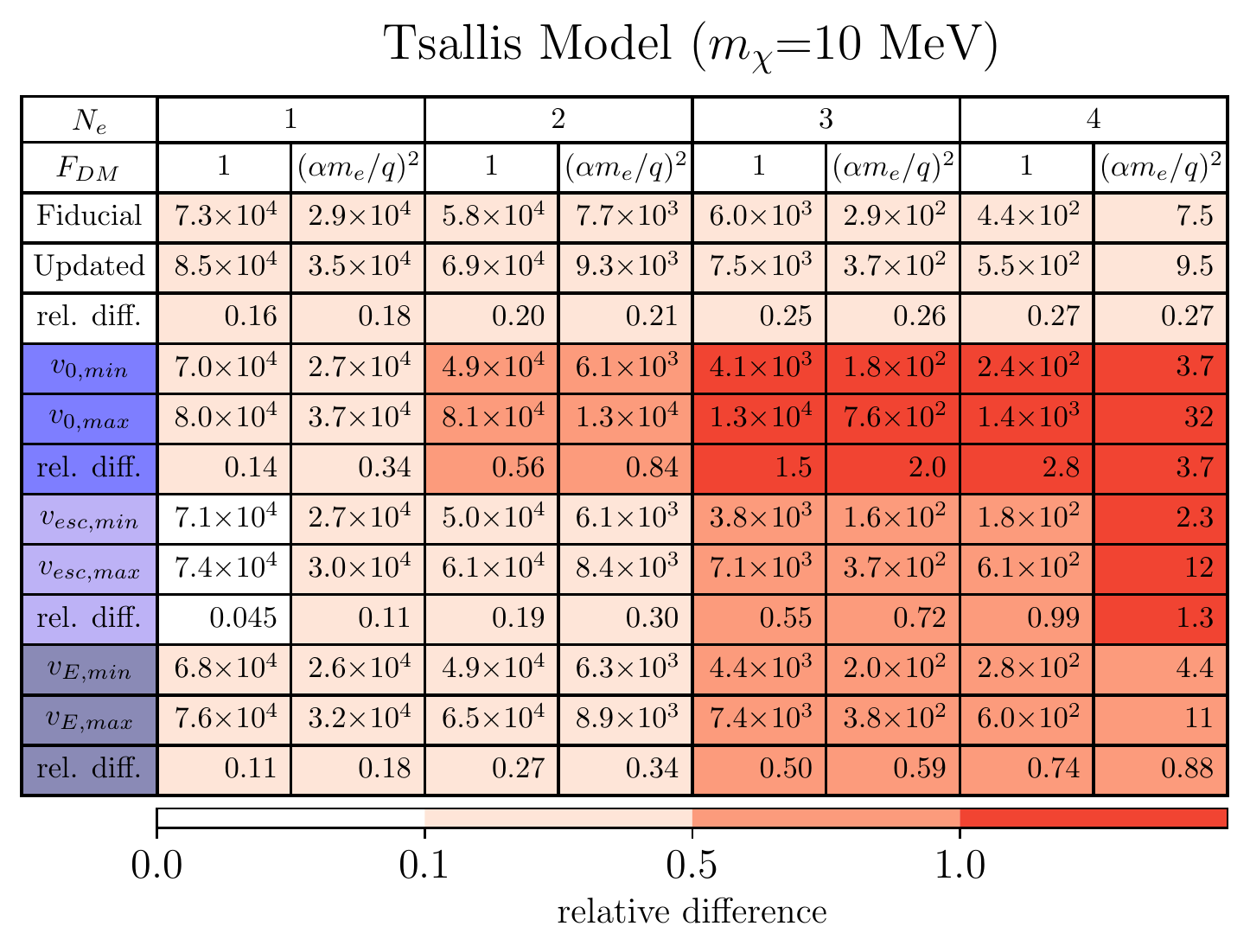}
\includegraphics[width=0.8\textwidth]{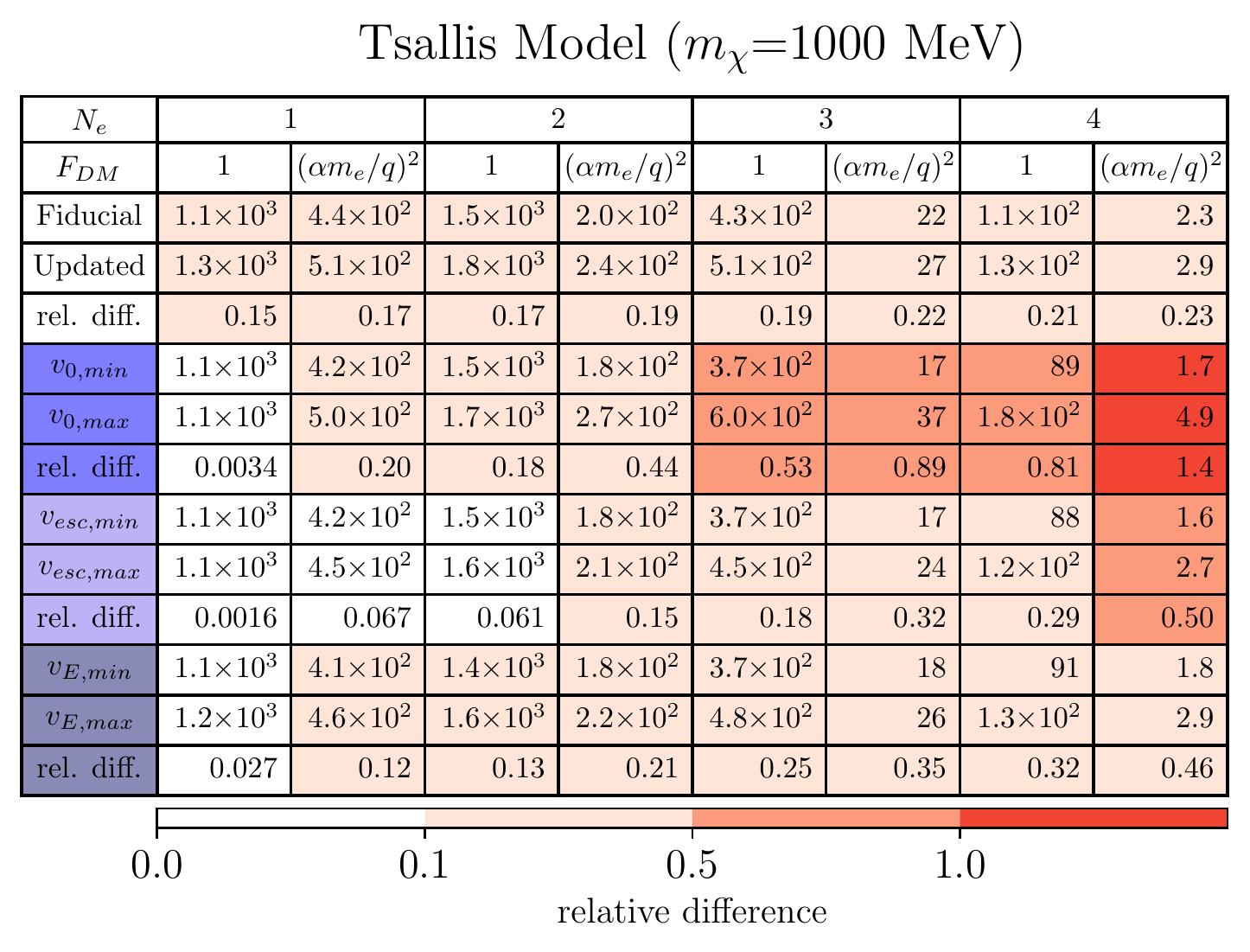}
    \caption{Expected number of events in 1 kg-year for various values of the Tsallis parameters $v_0, \vesc$, and $v_E$, as well as the relative difference between the minimum and maximum values for each parameter, for a given $N_e$ bin and DM form factor $F_{\rm DM}$. The top panel displays the results for $m_\chi=10$ MeV, and the bottom panel for $m_\chi=1$ GeV. The color scale indicates the size of the relative difference, with white (red) being the smallest (largest) relative difference.}
    \label{tab:Tsarate}
\end{table}
\newpage
\FloatBarrier
\begin{table}[H]
    \centering
\includegraphics[width=0.8\textwidth]{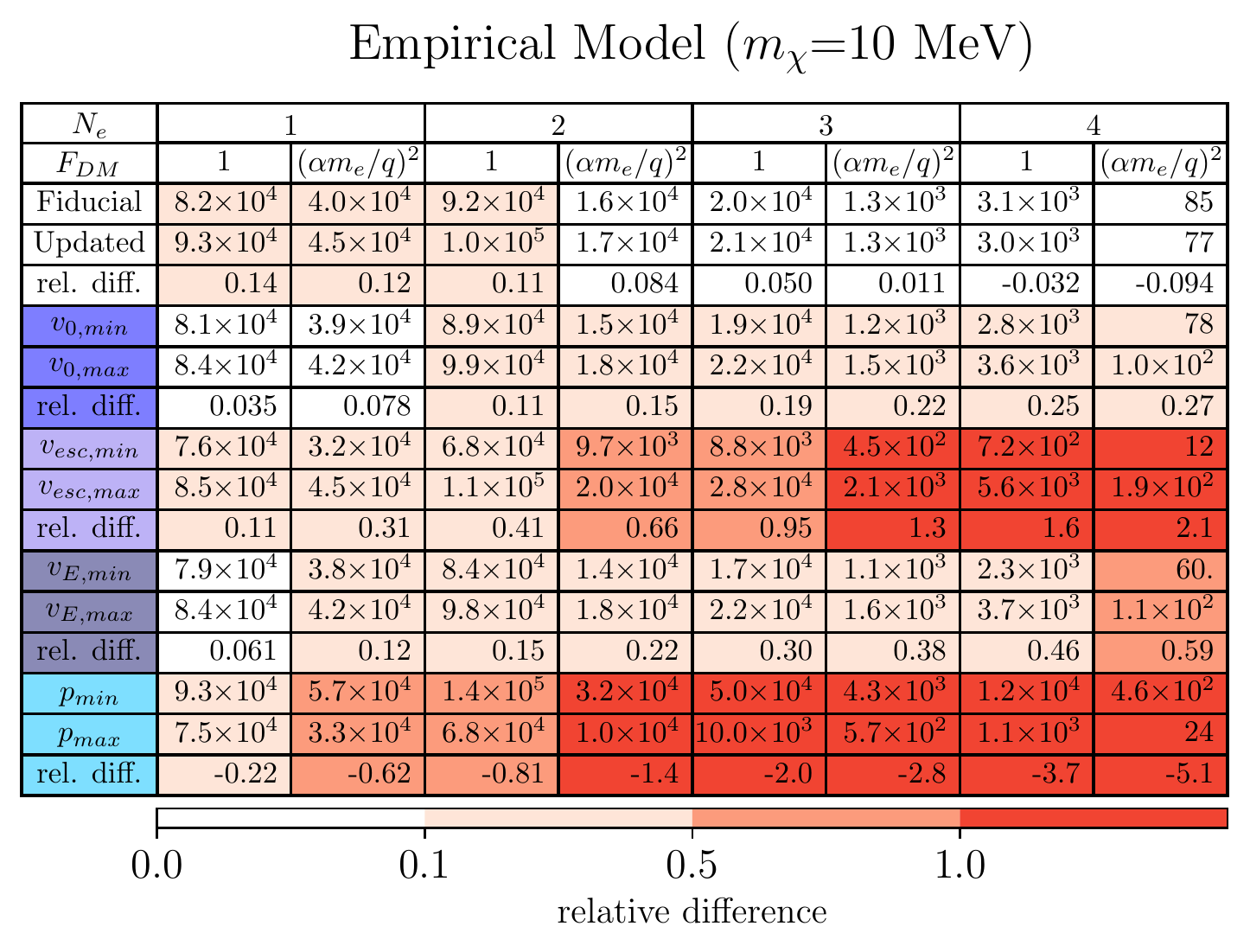}
\includegraphics[width=0.8\textwidth]{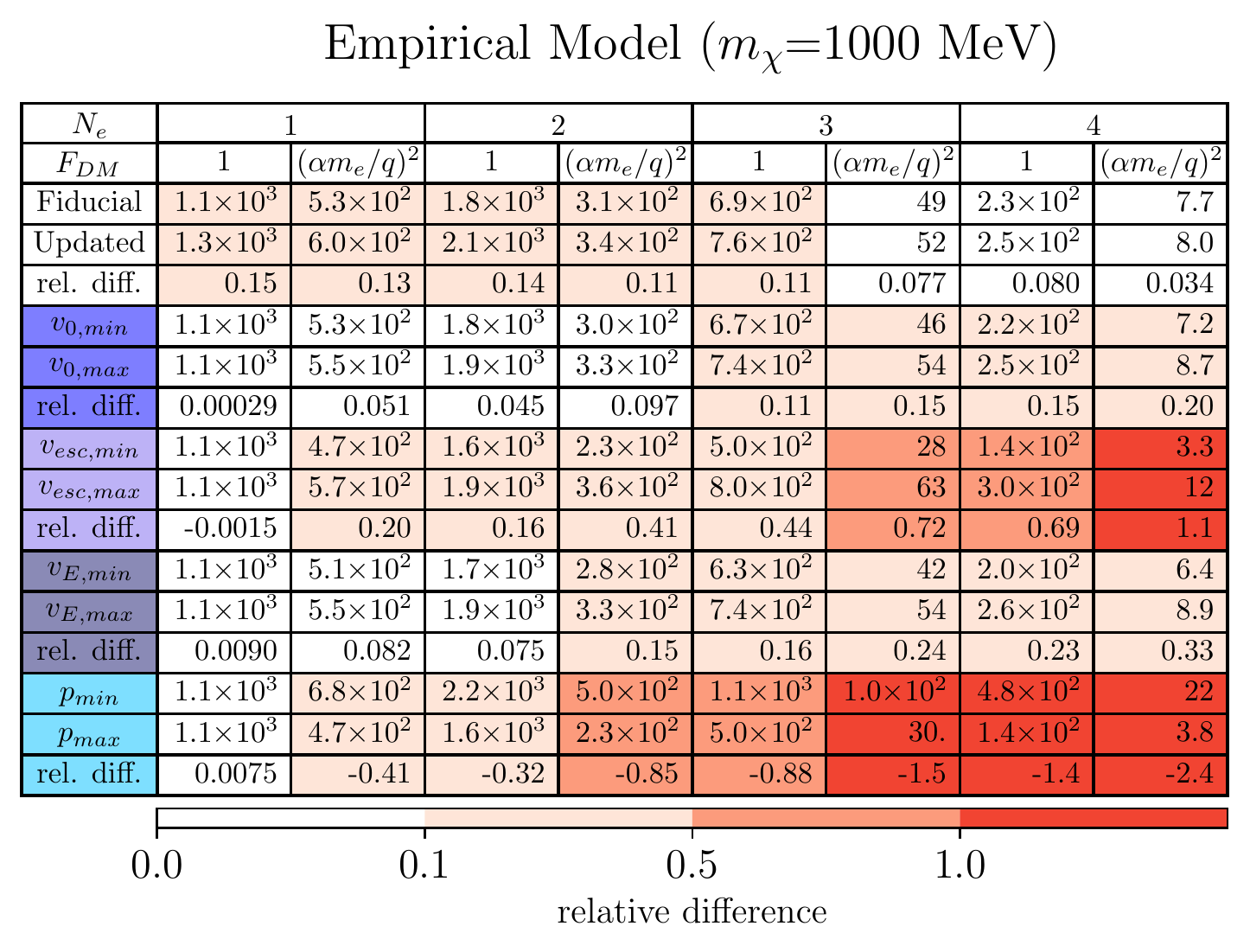}
    \caption{Expected number of events in 1 kg-year for various values of the empirical parameters $v_0, \vesc, v_E$, and $p$, as well as the relative difference between the minimum and maximum values for each parameter, for a given $N_e$ bin and DM form factor $F_{\rm DM}$. The top panel displays the results for $m_\chi=10$ MeV, and the bottom panel for $m_\chi=1$ GeV. The color scale indicates the size of the relative difference, with white (red) being the smallest (largest) relative difference.}
    \label{tab:MSWrate}
\end{table}
\end{appendices}

\newpage
\bibliographystyle{JHEP}
\bibliography{ref}
\end{document}